\documentclass[aps,prc,floatfix,nofootinbib,showpacs,twocolumn,superscriptaddress]{revtex4-1}

\usepackage{mathrsfs}
\usepackage{amsmath}
\usepackage{amssymb}
\usepackage{bbold}
\usepackage{color}
\usepackage{graphicx}
\usepackage{soul}

\usepackage[mathscr,scaled=1.15]{urwchancal}
\DeclareFontFamily{OT1}{pzc}{}
\DeclareFontShape{OT1}{pzc}{m}{it}%
{<-> s * [1.15] pzcmi7t}{}
\DeclareMathAlphabet{\mathpzc}{OT1}{pzc}{m}{it}

\definecolor{purple}{rgb}{0.5,0,0.5}
\definecolor{blue}{rgb}{0.0,0,0.9}

\begin{document}

%%%--preprint numbers
%\vspace*{-5.8ex}
%\hspace*{\fill}{NPAC-12-11}

%\vspace*{+4.8ex}

\title{Nucleon tensor charges and electric dipole moments}

\author{Mario Pitschmann}
\affiliation{Atominstitut, Technische Universit\"at Wien, Stadionallee 2, A-1020 Wien, Austria}

\author{Chien-Yeah Seng}
\affiliation{Amherst Center for Fundamental Interactions, Department of Physics,
University of Massachusetts Amherst, Amherst, MA 01003, USA}

\author{Craig D.\ Roberts}
\affiliation{Physics Division, Argonne National Laboratory, Argonne, Illinois 60439, USA}

\author{Sebastian M.~Schmidt}
\affiliation{Institute for Advanced Simulation, Forschungszentrum J\"ulich and JARA, D-52425 J\"ulich, Germany}

\date{7 November 2014}
%\date{1 September 2014}
%\date{04 August 2014}

\begin{abstract}
A symmetry-preserving Dyson-Schwinger equation treatment of a vector-vector contact interaction is used to compute dressed-quark-core contributions to the nucleon $\sigma$-term and tensor charges.  The latter enable one to directly determine the effect of dressed-quark electric dipole moments (EDMs) on neutron and proton EDMs.  The presence of strong scalar and axial-vector diquark correlations within ground-state baryons is a prediction of this approach.  These correlations are active participants in all scattering events and thereby modify the contribution of the singly-represented valence-quark relative to that of the doubly-represented quark.  Regarding the proton $\sigma$-term and that part of the proton mass which owes to explicit chiral symmetry breaking, with a realistic $d$-$u$ mass splitting the singly-represented $d$-quark contributes 37\% more than the doubly-represented $u$-quark; and in connection with the proton's tensor charges, $\delta_T u$, $\delta_T d$, the ratio $\delta_T d/\delta_T u$ is 18\% larger than anticipated from simple quark models.  Of particular note, the size of $\delta_T u$ is a sensitive measure of the strength of dynamical chiral symmetry breaking; and $\delta_T d$ measures the amount of axial-vector diquark correlation within the proton, vanishing if such correlations are absent.
$\,$\\[2ex]\hspace*{\fill}{\emph{Preprint no}. ACFI-T14-23}\\[-8ex]
\end{abstract}

\pacs{
12.38.Lg,   %	Other nonperturbative calculations
14.20.Dh,   %	Protons and neutrons
13.88.+e,   %	Polarization in interactions and scattering
11.30.Er	% Charge conjugation, parity, time reversal, and other discrete symmetries
}

\maketitle

\section{Introduction}
%\noindent \emph{\textbf{Background}}.
In recent years a global approach to the description of nucleon structure has emerged, one in which we may express our knowledge of the nucleon in the Wigner distributions of its basic constituents and thereby provide a multidimensional generalisation of the familiar parton distribution functions (PDFs).  The Wigner distribution is a quantum mechanics concept analogous to the classical notion of a phase space distribution.  Following from such distributions, a natural interpretation of measured observables is provided by construction of quantities known as generalised parton distributions (GPDs) \cite{Dittes:1988xz,Ji:1996nm,Radyushkin:1996nd,Mueller:1998fv,Goeke:2001tz,Diehl:2003ny,%
Belitsky:2005qn,Boffi:2007yc} and transverse momentum-dependent distributions (TMDs) \cite{Ralston:1979ys,Sivers:1989cc,Kotzinian:1994dv,Mulders:1995dh,%
Collins:2003fm,Belitsky:2003nz,Bacchetta:2006tn}: GPDs are linked to a \textit{spatial} tomography of the nucleon; and TMDs allow for its \textit{momentum} tomography.  A new generation of experiments aims to provide the empirical information necessary to develop a phenomenology of nucleon Wigner distributions.  %However, this will provide little information about the Standard Model's strong interaction sector unless these distributions can be calculated in a framework with a well-defined connection to QCD.

\begin{figure}[b!]
\begin{center}
\includegraphics[clip,width=0.3\textwidth]{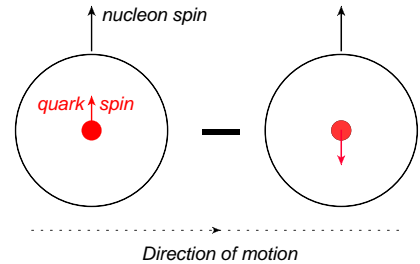}
\caption{\label{figTensorCharge} The tensor charge, Eq.\,\eqref{DefineTensorCharge}, measures the net light-front distribution of transversely polarised quarks inside a transversely polarized proton.}
\end{center}
\end{figure}

At leading-twist there are eight distinct TMDs, only three of which are nonzero in the collinear limit; i.e., in the absence of parton transverse momentum within the target, $k_\perp = 0$: the unpolarized $(f_1)$, helicity $(g_{1L})$ and transversity $(h_{1T})$ distributions.  In connection with the last of these, one may define the proton's tensor charges ($q=u,d,\ldots$)
\begin{equation}
\label{DefineTensorCharge}
\delta_T q = \int_{-1}^1 dx\, h^q_{1T}(x) = \int_0^1 dx\, \left[ h_{1T}^q(x) - h_{1T}^{\bar q}(x)\right]\,,
\end{equation}
which, as illustrated in Fig.\,\ref{figTensorCharge}, measures the light-front number-density of quarks with transverse polarisation parallel to that of the proton minus that of quarks with antiparallel polarisation; viz., it measures any bias in quark transverse polarisation induced by a polarisation of the parent proton.  The charges $\delta_T q$ represent a close analogue of the nucleon's flavour-separated axial-charges, which measure the difference between the light-front number-density of quarks with helicity parallel to that of the proton and the density of quarks with helicity antiparallel \cite{Chang:2012cc}.  In nonrelativistic systems the helicity and transversity distributions are identical because boosts and rotations commute with the Hamiltonian.

The transversity distribution is measurable using Drell-Yan processes in which at least one of the two colliding particles is transversely polarised \cite{Barone:2001sp}; but such data is not yet available.  Alternatively, the transversity distribution is accessible via semi-inclusive deep-inelastic scattering using transversely polarised targets and also in unpolarised $e^+ e^-$ processes, by studying azimuthal correlations between produced hadrons that appear in opposing jets ($e^+ e^- \to h_1 h_2 X$).  Capitalising on these observations, the transversity distributions were extracted through an analysis of combined data from the HERMES, COMPASS and Belle collaborations \cite{Anselmino:2013vqa}; and those distributions have been used to produce an estimate of the proton's tensor charges, with the following flavour-separated results:
\begin{equation}
\label{tensorud}
%red-circles
\delta_T u = 0.39 ^{+0.18}_{-0.12} \,,\quad
\delta_T d = -0.25 ^{+0.30}_{-0.10} \,,
%or red-triangles
%\delta_T u = 0.31 ^{+0.16}_{-0.12} \,,\quad
%\delta_T d = -0.27 ^{+0.10}_{-0.10} \,,
\end{equation}
at a renormalisation scale $\zeta_A = 0.9\,$GeV.  Given that the tensor charges are a defining intrinsic property of the nucleon, the magnitude of the errors in Eqs.\,\eqref{tensorud} is unsatisfactory.  It is therefore critical to better determine $\delta_T u $, $\delta_T d$.  Consequently, following upgrades at the Thomas Jefferson National Accelerator Facility (JLab), it is anticipated \cite{Dudek:2012vr} that experiments \cite{Gao:2010av,Avakian:2014aba} in Hall-A (SoLID) and Hall-B (CLAS12) will provide a far more precise determination of the tensor charges.
%It will be measured in Hall B using CLAS [3-14, 3-55] and in Hall A using SoLID [3-17, 3-18] and will be determined with much improved precision in the 12 GeV era.

Naturally, measurement of the transversity distribution and the tensor charges will not reveal much about the strong interaction sector of the Standard Model unless these quantities can be calculated using a framework with a traceable connection to QCD.  This point is emphasised with particular force by the circumstances surrounding the pion's valence-quark PDF.  As reviewed elsewhere \cite{Holt:2010vj}, numerous authors suggested that QCD was challenged by a PDF parametrisation based on a precise $\pi N$ Drell-Yan measurement \cite{Conway:1989fs}.  However, the appearance of nonperturbative calculations within the framework of continuum QCD \cite{Hecht:2000xa,Nguyen:2011jy} forced reanalyses of the cross-section, with the inclusion of next-to-leading-order evolution \cite{Wijesooriya:2005ir} and soft-gluon resummation \cite{Aicher:2010cb}, so that now those claims are known to be false and the pion's valence-quark PDF may be viewed as a success for QCD \cite{Chang:2014lva}.  The comparisons between experiment and computations of the pion and kaon parton distribution amplitudes and electromagnetic form factors have reached a similar level of understanding \cite{Chang:2013nia,Shi:2014uwa}.

Herein, therefore, we compute the proton tensor charges using a confining, symmetry-preserving Dyson-Schwinger equation (DSE) treatment of a single quark-quark interaction; namely, a vector$\,\otimes\,$vector contact-interaction.  This approach has proved useful in a variety of contexts, which include meson and baryon spectra, and their electroweak elastic and transition form factors
\cite{GutierrezGuerrero:2010md,Roberts:2010rn,Roberts:2011wy,Roberts:2011cf,Wilson:2011aa,%
Chen:2012qr,Chen:2012txa,Pitschmann:2012by,Wang:2013wk,Segovia:2013rca,Segovia:2013uga}. In fact, so long as the momentum of the probe is smaller in magnitude than the dressed-quark mass produced by dynamical chiral symmetry breaking (DCSB), many results obtained in this way are practically indistinguishable from those produced by the most sophisticated interactions that have thus far been employed in DSE studies \cite{Roberts:2007jh,Chang:2011vu,Bashir:2012fs,Cloet:2013jya}.

It is apposite to remark here that confinement and DCSB are two key features of QCD; and much of the success of the contact-interaction approach owes to its efficacious expression of these emergent phenomena in the Standard Model.  They are explained in some detail elsewhere \cite{Roberts:2007jh,Chang:2011vu,Bashir:2012fs,Cloet:2013jya} so that here we remark only that confinement may be expressed via dynamically-driven changes in the analytic structure of QCD's propagators and
vertices; and DCSB is the origin of more than 98\% of the mass of visible material in the Universe.  These phenomena are intimately connected; and whereas the nature of confinement is still debated, DCSB is a theoretically established nonperturbative feature of QCD \cite{national2012Nuclear}, which has widespread, measurable impacts on hadron observables, e.g. Refs.\,\cite{Chen:2012qr,Pitschmann:2012by,Chang:2013pq,Cloet:2013gva,Roberts:2013mja,Chang:2013epa,%
Gao:2014bca,Shi:2014uwa,Segovia:2014aza}, so that its expression in QCD is empirically verifiable.

Apart from the hadron physics imperative, the value of the nucleon tensor charges can be directly related to the visible impact of a dressed-quark electric dipole moment (EDM) on neutron and proton EDMs \cite{Hecht:2001ry}.  Novel beyond-the-Standard-Model (BSM) scalar operators may also conceivably be measurable in precision neutron experiments so that one typically considers both the nucleon scalar and tensor charges when exploring bounds on BSM physics \cite{Bhattacharya:2011qm}.  The sum of the scalar charges of all active quark flavours is simply the nucleon $\sigma$-term, which we therefore also compute herein.

Relying on material provided in numerous appendices, we provide a brief outline of our computational framework in Sec.\,\ref{framework}: both the Faddeev equation treatment of the nucleon and the currents which describe the interaction of a probe with a baryon composed from consistently-dressed constituents.  This presentation scheme enables us to embark quickly upon the description and analysis of our results for the scalar and tensor charges, Secs.\,\ref{SecSigma} and \ref{SecTensor}, respectively.  In Sec.\,\ref{SecEDM} we use our results for the tensor charges in order to determine the impact of valence-quark EDMs on the neutron and proton EDMs.  Section~\ref{SecConclusion} is an epilogue.

\label{secCharges}
\section{Nucleon Faddeev Amplitude and Relevant Interaction Currents}
\label{framework}
\begin{figure}[t]
\centerline{%
\includegraphics[clip,width=0.4\textwidth]{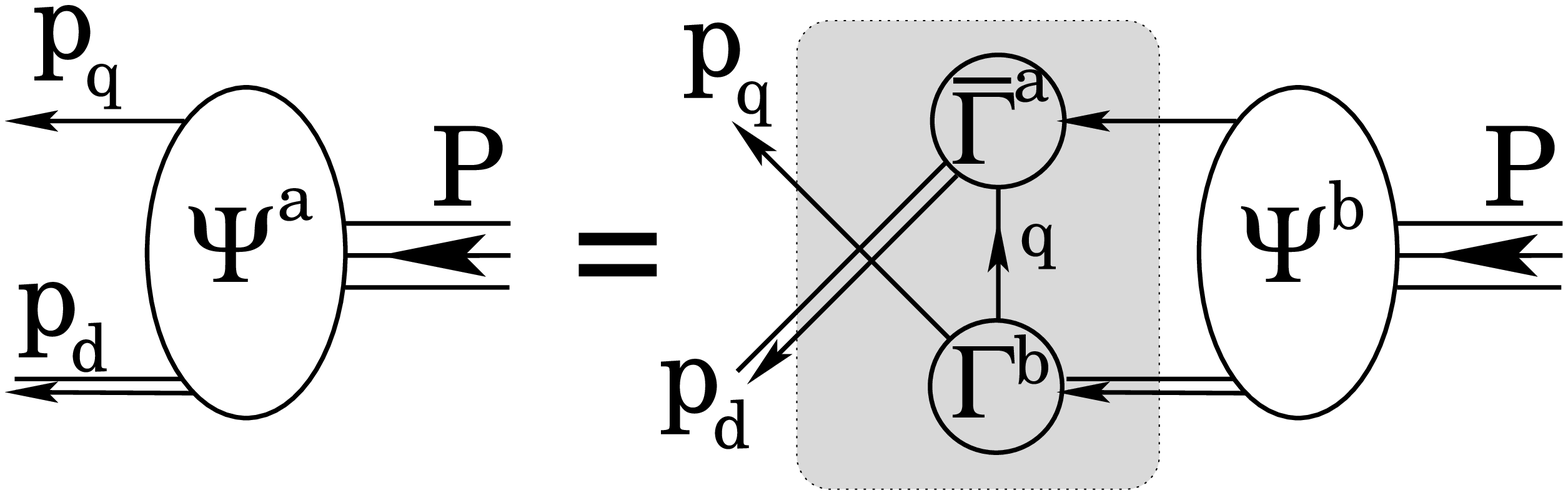}}
\caption{\label{figFaddeev} Poincar\'e covariant Faddeev equation.  $\Psi$ is the Faddeev amplitude for a nucleon of total momentum $P= p_q + p_d$.  The shaded rectangle demarcates the kernel of the Faddeev equation: \emph{single line}, dressed-quark propagator; $\Gamma$, diquark correlation (Bethe-Salpeter) amplitude; and \emph{double line}, diquark propagator.  (See Apps.\,\ref{sec:contact}, \ref{sec:Faddeev} for details.)}
\end{figure}

We base our description of the nucleon's dressed-quark-core on solutions of a Faddeev equation, which is illustrated in Fig.\,\ref{figFaddeev}, and formulated and described in Apps.\,\ref{sec:contact}, \ref{sec:Faddeev}.  In order to determine the scalar and tensor charges of the nucleon described by this Faddeev equation, the $Q^2=0$ values of three interaction currents are needed: elastic electromagnetic, which determines the canonical normalisation of the nucleon's Faddeev amplitude; scalar; and tensor.  The computation of these quantities is detailed in App.\,\ref{ICurrents}.

\section{Sigma-Term}
\label{SecSigma}
The contribution of a given quark flavour ($q=u,d,\ldots$) to a nucleon's $\sigma$-term is defined by the matrix element
\begin{equation}
\sigma_q = m_q \, \langle N(p) | \bar q {\mathbb 1} q | N(p)\rangle\,,
\end{equation}
where $|N(p)\rangle$ is the state vector of a nucleon with four-momentum $p$.  The $\sigma$-term is independent of the renormalisation scale used in the computation, even though the individual pieces in the product on the right-hand-side (rhs) are not.  As explained in App.\,\ref{AppModelScale}, the scale appropriate to our symmetry-preserving regularisation of the contact interaction is $\zeta_H \approx M$, where $M$ is the dressed-quark mass.

Our computed value of the nucleon's $\sigma$-term is reported in Eq.\,\eqref{eqNsigma}; viz.,
\begin{equation}
\label{eqNsigmaF}
\sigma_N = \sigma_u + \sigma_d = m \, 3.05 = 21\,{\rm MeV}.
\end{equation}
%%%-This is 10% smaller than obtained via the Feynman-Hellman theorem: 26-27 MeV
This result is consistent with that obtained using the Feynman-Hellmann theorem in connection with the results from which Ref.\,\cite{Roberts:2011cf} was prepared.  An interesting way to expose this is to recall Eq.\,\eqref{NucleonEigenVector}, which states that our analysis describes a nucleon that is 77\% dressed-quark$\,+\,$scalar-diquark and 23\% dressed-quark$\,+\,$axial-vector diquark.  In the isospin symmetric limit, which we typically employ, it follows that
\begin{eqnarray}
\sigma_N &=& 0.77 \left[ \sigma_Q + \sigma_{qq^0} \right] + 0.23 \left[ \sigma_Q + \sigma_{qq^1} \right] \\
&=& \sigma_Q + 0.77\, \sigma_{qq^0} + 0.23 \, \sigma_{qq^1},
\label{eqSigmaNSum}
\end{eqnarray}
where
\begin{subequations}
\label{constituentsigmaterms}
\begin{eqnarray}
\sigma_Q &=&  m \frac{\partial M}{\partial m} = 9.6\,{\rm MeV},\\
\sigma_{qq^0} &=&  m \frac{\partial m_{qq^0}}{\partial m} = 16\,{\rm MeV},\\
\sigma_{qq^1} &=&  m \frac{\partial m_{qq^1}}{\partial m} = 10\,{\rm MeV},
\end{eqnarray}
\end{subequations}
again computed using material in Ref.\,\cite{Roberts:2011cf}.  Inserting Eqs.\,\eqref{constituentsigmaterms} into Eq.\,\eqref{eqSigmaNSum}, one obtains $\sigma_N=24\,$MeV.\footnote{The origin of the 11\% mismatch is explained in Sec.\,\ref{sscc}.}  Apparently, so far as the contribution of explicit chiral symmetry breaking to the mass of the nucleon's dressed-quark core is concerned, the contact-interaction nucleon is a simple system.  This analysis also shows that our diagrammatic computational method is sound; and hence Eq.\,\eqref{eqNsigmaF} is the rainbow-ladder (RL) truncation\footnote{The rainbow-ladder truncation is the leading-order term in the most widely used, global-symmetry-preserving DSE truncation scheme \cite{Munczek:1994zz,Bender:1996bb}.}
prediction of a vector$\,\otimes\,$vector contact-interaction treated in the Faddeev equation via the static approximation.  (Inclusion of meson-baryon loop effects will increase the result in Eq.\,\eqref{eqNsigmaF} by approximately 15\% \cite{Flambaum:2005kc}.)

In addition, the fact that Eqs.\,\eqref{eqNsigmaF} and \eqref{eqSigmaNSum} yield similar results emphasises the important role of diquark correlations because if the nucleon were just a sum of three massive, weakly-interacting dressed-quarks, then one would have
\begin{equation}
\sigma_N^{3M} = 3 \, \sigma_Q = 29\,{\rm MeV}\,,
\end{equation}
which is 21\% too large.
% cf. FHT in Eq.(6).
%... not useful comparison ... It is notable that our computed value of $\sigma_N$ is also commensurate with the value $28\,$MeV obtained using propagators and vertices with QCD-like momentum dependence in a nucleon constituted from a scalar-diquark alone \cite{Bloch:1999rm}.

Adopting a different perspective, we note that the value in Eq.\,\eqref{eqNsigmaF} is roughly one-half that produced by a Faddeev equation kernel that incorporates scalar and axial-vector diquark correlations in addition to propagators and interaction vertices that possess QCD-like momentum dependence \cite{Flambaum:2005kc}.  It compares similarly with the value inferred in a recent analysis \cite{Shanahan:2012wh} of lattice-QCD results for octet baryon masses in $2+1$-flavour QCD:
\begin{equation}
\sigma_N = 45\pm 6 \,{\rm MeV}\,.
\end{equation}

In order to understand the discrepancy, consider Eqs.\,\eqref{constituentsigmaterms}.  The value of $\sigma_Q$ matches expectations based on gap equation kernels whose ultraviolet behaviour is consistent with QCD \cite{Flambaum:2005kc,Maris:1999bh}.  On the other hand, with such interactions one typically finds $\sigma_{qq^0} \gtrsim \sigma_{qq^1} \gtrsim \sigma_{\rho} = 25\,$MeV.  We therefore judge that Eq.\,\eqref{eqNsigmaF} underestimates the physical value of $\sigma_N$; and that the mismatch originates primarily in the rigidity of the diquark Bethe-Salpeter amplitudes produced by the contact interaction, which leads to weaker $m$-dependence of the diquark (and hence nucleon) masses than is obtained with more realistic kernels.\footnote{Consider that if one uses $\sigma_{qq^0}=\sigma_{qq^1}=30\,$MeV, then $\sigma_N\approx 40\,$MeV.}
% s_Q + 0.77 s0 + 0.23 s1
% s0->30 s1->25 sN->36
Notwithstanding this, Eq.\,\eqref{eqNsigmaF} is a useful benchmark, providing a sensible result via a transparent method.

%of both the diquark Bethe-Salpeter amplitudes produced by the contact interaction and the nucleon Faddeev amplitudes obtained using the so-called static approximation, Eq.\,\eqref{staticexchange}.  The latter two properties lead to weaker $m$-dependence of the diquark and nucleon masses than is obtained with more realistic interactions.  The simplicity of our Faddeev equation kernel also eliminates two-loop contributions to the scalar-nucleon current.  Notwithstanding this, Eq.\,\eqref{eqNsigmaF} is a useful benchmark, providing a sensible result via a transparent method.

Further valuable information may be obtained from the results in App.\,\ref{AppScalarCurrent} if one supposes that the ratio of contact-interaction $d$- and $u$-quark contributions is more reliable than the net value of $\sigma_N$.  In this connection, note that for a proton constituted as a weakly interacting system of three massive dressed-quarks in the isospin symmetric limit
\begin{equation}
\frac{\sigma_{N,d}^{3M}}{\sigma_{N,u}^{3M}} = \frac{1}{2}\,.
\end{equation}
Comparing this with our computed value
\begin{equation}
\frac{\sigma_{N,d}}{\sigma_{N,u}} = 0.65\,,
%= 0.75\,, ... unrescaled
\end{equation}
one learns that diquark correlations work to accentuate the contribution of the singly-represented valence-quark to the proton $\sigma$-term relative to that of doubly-represented valence-quarks: the magnification factor is $1.3$.

Let's take this another step and assume that $\hat\sigma_{N,u}$, $\hat\sigma_{N,d}$ in App.\,\ref{AppScalarCurrent} respond weakly to changes in $m$.  This is valid so long as solutions of the dressed-quark gap equation satisfy
\begin{equation}
\left.\frac{dM}{dm}\right|_{(m_u+m_d)/2} \stackrel{m_u,m_d \ll M} {\approx} \left.\frac{dM}{dm}\right|_{m_u,m_d},
\end{equation}
which is found to be a good approximation in all available studies (see, e.g., Refs.\,\cite{Holl:2005st,Pennington:2010gy}).
%\footnote{With a momentum-dependent interaction, this behaviour is required (and found) on $k^2 \lesssim m_N^2$; see, e.g., Ref.\,\cite{Pennington:2010gy}.}
One may then estimate the effects of isospin symmetry violation owing to the difference between $u$- and $d$-quark current-masses. Taking the value of the mass ratio from Ref.\,\cite{Beringer:1900zz}, one finds
\begin{equation}
\label{split1}
\frac{m_u}{m_d} =0.48 \pm 0.1 \quad \Rightarrow \quad
\frac{m_d \, \hat \sigma_{N,d}}{m_u \, \hat \sigma_{N,u}} = 1.35^{+0.47}_{-0.30}\,.
\end{equation}
%mu=2.3
%md=4.8
%mu/md=0.48 +/-0.1
Alternatively, one might use the mass ratio inferred from a survey of numerical simulations of lattice-regularised QCD \cite{Colangelo:2010et}, in which case
\begin{equation}
\label{FLAGresult}
\frac{m_u}{m_d} =0.47 \pm 0.04 \quad \Rightarrow \quad \frac{m_d \, \hat \sigma_{N,d}}{m_u \,\hat \sigma_{N,u}} = 1.38^{+0.17}_{-0.14}\,.
\end{equation}
We predict, therefore, that the $d$-quark contribution to that part of the proton's mass which arises from explicit chiral symmetry breaking is roughly 37\% greater than that from the $u$-quark.  This value is commensurate with a contemporaneous estimate based on lattice-QCD \cite{Erben:2014hza}.  It is noteworthy that if the proton were a weakly interacting system of three massive dressed-quarks, then Eq.\,\eqref{FLAGresult} would yield $1.06^{+0.13}_{-0.11}$; and hence one finds again that the presence of diquark correlations within the proton enhances the contribution of $d$-quarks to the proton's $\sigma$-term.

\section{Tensor Charge}
\label{SecTensor}
The tensor charge associated with a given quark flavour in the proton is defined via the matrix element ($q = u,d,\ldots$)
\begin{align}\label{tcd}
\langle P(p,\sigma)|\bar q\sigma_{\mu\nu}q|P(p,\sigma)\rangle=\delta_T q\,\bar
{\mathpzc u}(p,\sigma)\sigma_{\mu\nu}{\mathpzc u}(p,\sigma) \,,
\end{align}
where ${\mathpzc u}(p,\sigma)$ is a spinor and $|P(p,\sigma)\rangle$ is a state vector describing a proton with momentum $p$ and spin $\sigma$.\footnote{In the isospin symmetric limit: $\delta_T^p u :=\delta_T u = \delta_T^n d$, $\delta_T^p d :=\delta_T d = \delta_T^n u$.}
With $\delta_T u$, $\delta_T d$ in hand, the isoscalar and isovector tensor charges are readily computed:
\begin{equation}
  g_T^{(0)} = \delta_T u + \delta_T d\,, \;
  g_T^{(1)} = \delta_T u - \delta_T d\,.
\end{equation}
Importantly, the tensor charge is a scale-dependent quantity.  Its evolution is discussed in App.\,\ref{AppEvolution}.

Our analysis of the proton's tensor charge in a symmetry-preserving RL-truncation treatment of a vector$\,\otimes\,$vector contact-interaction is detailed in App.\,\ref{AppTensor}.  At the model scale, $\zeta_H$, which is determined and explained in App.\,\ref{AppModelScale}, we obtain the results in Table~\ref{TT}, which represent a parameter-free prediction: the current-quark mass and the two parameters that define the interaction were fixed elsewhere \cite{Roberts:2011wy}, in a study of $\pi$- and $\rho$-meson properties.

It is natural to ask for an estimate of the systematic error in the values reported in Table~\ref{TT}.  As we saw in Sec.\,\ref{SecSigma}, the error might pessimistically be as much as a factor of two.  However, that is an extreme case because, as observed in the Introduction, one generally finds that our treatment of the contact interaction produces results for low-momentum-transfer observables that are practically indistinguishable from those produced by RL studies that employ more sophisticated interactions
\cite{GutierrezGuerrero:2010md,Roberts:2010rn,Roberts:2011cf,Roberts:2011wy,Wilson:2011aa,%
Chen:2012qr,Chen:2012txa,Pitschmann:2012by,Wang:2013wk,Segovia:2013rca,Segovia:2013uga}. It is therefore notable that analyses of hadron physics observables using the RL truncation and one-loop QCD renormalisation-group-improved (RGI) kernels for the gap and bound-state equations produce results that are typically within 15\% of the experimental value \cite{Roberts:2007jh}.  We therefore ascribe a relative error of 15\% to the results in Table~\ref{TT} so that our predictions are:
\begin{equation}
\label{errorResults}
\begin{array}{l|cccc}
&\delta_T u & \delta_T d & g_T^{(0)} & g_T^{(1)}\\
\zeta_H \approx M & 0.69(10) & -0.14(2) & 0.55(8) & 0.83(12)
\end{array}\,.
\end{equation}

One means by which to check our error estimate is to repeat the calculations described herein using a modern RGI kernel \cite{Qin:2011dd} in the gap and bound-state equations.  That has not yet been done but one may nevertheless infer what it might yield.  Consider first Ref.\,\cite{Yamanaka:2013zoa}, which computes the dressed-quark-tensor vertex using a RL-treatment of a QCD-based kernel: one observes that the dressed-quark's tensor charge is markedly suppressed; namely, with a QCD-based momentum-dependent kernel, a factor of approximately one-half appears on the rhs of Eq.\,\eqref{DressedQT}.  This DCSB-induced suppression would tend to reduce the values in Eq.\,\eqref{errorResults}.  On the other hand, the use of a more sophisticated momentum-dependent kernel in the bound-state equations increases the amount of dressed-quark orbital angular momentum in the proton, an effect apparent in the reduction of the fraction of proton helicity carried by dressed $u$- and $d$-quarks when one shifts from a contact-interaction framework to a QCD-kindred approach \cite{Roberts:2013mja,Segovia:2014aza}.  Hence, the tensor charges are determined by two competing effects, the precise balance amongst which can only be revealed by detailed calculations.

In this context, however, it is worth noting that similar DCSB-induced effects are observed in connection with $g_A$, the nucleon's axial charge.  The axial-charge of a dressed-quark is suppressed \cite{Chang:2012cc}, owing to DCSB; but that is compensated in the calculation of $g_A$ by dressed-quark orbital angular momentum in the nucleon's Faddeev wave-function, with the computed value of the nucleon's axial-charge being 20\% larger than that of a dressed-quark.  The net effect is that a computation of $g_A$ using the framework of Refs.\,\cite{Segovia:2014aza} can readily produce a result that is within 15\% of the empirical value \cite{Roberts:2007jh,Chang:2012cc}.  This suggests that our error estimate is reasonable.

The predictions in Eq.\,\eqref{errorResults} are quoted at the model scale, whose value is explained in App.\,\ref{AppModelScale}.  In order to make a sensible comparison with estimates obtained in modern simulations of lattice-regularised QCD, those results must be evolved to $\zeta_2=2\,$GeV.  We therefore list here the results obtained under leading-order evolution to $\zeta_2=2\,$GeV, obtained via multiplication by the factor in Eq.\,\eqref{Efactor}:
\begin{equation}
\label{tensorz2}
\begin{array}{l|cccc}
&\delta_T u & \delta_T d & g_T^{(0)} & g_T^{(1)}\\
\zeta_2 & 0.55(8) & -0.11(2) & 0.44(7) & 0.66(10)
\end{array}\,.
\end{equation}
The error in Eq.\,\eqref{Efactor} does not propagate significantly into these results.

%--Old before rescaling It is notable that the dominant contribution to $\delta_T u$ arises from Diagram~1, in which the tensor probe interacts with a dressed-quark while a scalar diquark is the bystander.  The sum of Diagrams~5 and 6, which describes the tensor probe causing a transition between scalar- and axial-vector diquark correlations within the proton whilst the dressed-quark is a bystander, provides the next most important contribution in magnitude.  It is a large negative contribution for both $\delta_T u$ and $\delta_T d$: indeed, owing to a near exact cancellation between Diagrams~2 and 4 in the $d$-quark sector, which describe the quark$\,+\,$axial-vector-diquark contributions, the sum of Diagrams~5 and 6 provides almost the entire result for $\delta_T d$.
%%
Notably, the dominant contribution to $\delta_T u$ arises from Diagram~1: tensor probe interacting with a dressed $u$-quark with a scalar diquark as the bystander.  The tensor probe interacting with the axial-vector diquark, with a dressed-quark as a spectator, Diagram~4, produces the next largest piece.  However, that is largely cancelled by the sum of Diagrams~5 and 6: tensor probe causing a transition between scalar- and axial-vector diquark correlations within the proton whilst the dressed-quark is a bystander.  It is a large negative contribution for both $\delta_T u$ and $\delta_T d$: indeed, owing to a significant cancellation between Diagrams~2 and 4 in the $d$-quark sector, which describe the net result from quark$\,+\,$axial-vector-diquark contributions, the sum of Diagrams~5 and 6 provides almost the entire result for $\delta_T d$.

\begin{figure}[t!]
\includegraphics[width=1.0\linewidth]{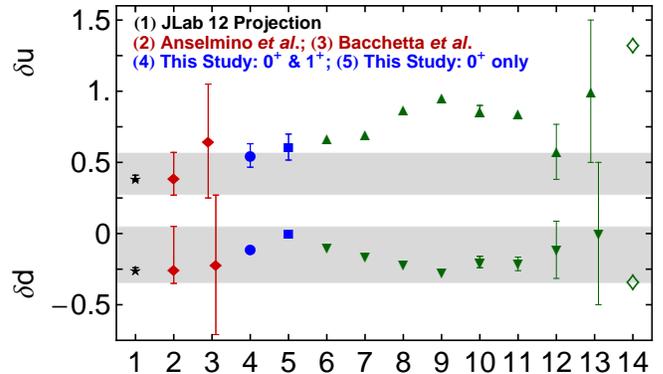}
\caption{Flavour separation of the proton's tensor charge:
``1'' -- illustration of anticipated accuracy in planned JLab experiment \cite{Gao:2010av}, with central values based on Eq.\,\eqref{tensorud};
``2'' -- results in Eq.\,\eqref{tensorud}, drawn from Ref.\,\cite{Anselmino:2013vqa};
``3'' phenomenological estimate in Ref.\,\cite{Bacchetta:2012ty}
``4'' -- prediction herein, Eq.\,\eqref{tensorz2};
``5'' -- result obtained herein with omission of axial-vector diquark correlations, Eq.\,\eqref{tensornoav};
``6-13'' -- estimates from Refs.\,\cite{Hecht:2001ry,Cloet:2007em,Pasquini:2006iv,Wakamatsu:2007nc,%
alexandrou:2014,Gockeler:2005cj,Gamberg:2001qc,He:1994gz}, respectively.
By way of context, we note that were the proton a weakly-interacting collection of three massive valence-quarks described by an SU$(4)$-symmetric spin-flavour wave function, then \cite{He:1994gz} the quark axial and tensor charges are identical, so that $\delta_T u =4/3$ and $\delta_T d=-1/3$ at the model scale.  These values are located at ``14''.
\label{TensorAll}}
\end{figure}

A particularly important result is the impact of the proton's axial-vector diquark correlation.  As determined in App.\,\ref{TensorScalarOnly}, with a symmetry-preserving treatment of a contact interaction, $\delta_T d$ is only nonzero if axial-vector diquark correlations are present.  Significantly, in dynamical calculations the strength of axial-vector diquark correlations relative to scalar diquark correlations is a measure of DCSB \cite{Chen:2012qr}.  In the absence of axial-vector diquark correlations [Eqs.\,\eqref{NoAxial}, Eq.\,\eqref{Efactor}]
\begin{equation}
\label{tensornoav}
\begin{array}{l|cccc}
& \delta_{T\!\not\, 1} u & \delta_{T\!\not\, 1} d &  g_{T\!\not\, 1}^{(0)} & g_{T\!\not\, 1}^{(1)} \\
\zeta_2 & 0.61(9) & 0 & 0.61(9) & 0.61(9)
\end{array}\,;
\end{equation}
i.e., $\delta_T d$ vanishes altogether and $\delta_T u$ is increased by 11\%.  We expect that the influence of axial-vector diquark correlations will be qualitatively similar in the treatment of more sophisticated kernels for the gap and bound-state equations.
A hint in support of this expectation may be drawn from the favourable comparison, depicted in Fig.\,\ref{TensorAll}, between the predictions for $\delta_T u$ in Eq.\,\eqref{tensornoav}, ``4'', and the result of Ref.\,\cite{Hecht:2001ry}, ``5''.  The latter employed a proton and tensor-current that suppressed but did not entirely eliminate the contribution from axial-vector diquark correlations.
%error estimate realistic ... emphasised by agreement of no.1+ with hecht ... suppressed but did not completely eliminate 1+ contributions to tensor charge owing to structure of the current
This same comparison also supports the verity of our error estimate.

Additionally, it is valuable to note that the magnitude of $\delta_T u$ is a direct probe of the strength of DCSB and hence of the strong interaction at infrared momenta.  This could be anticipated, e.g., from Eqs.\,\eqref{D1Tensor}, \eqref{D4Tensor}, the expressions for Diagrams~1 and 4, which produce the dominant positive contributions to $\delta_T u$: both show a strong numerator dependence on the dressed-quark mass, $M$; and $M/m\gg 1$ is a definitive signal of DCSB.  To quantify the effect, we reduced $\alpha_{\rm IR}$ in the gap and Bethe-Salpeter equations by 20\% and recomputed all relevant quantities.  This modification reduced the dressed-quark mass by 33\%: $M = 0.368 \to M_< = 0.246\,$GeV.  Combined with knock-on effects throughout all correlations and bound-states, the 20\% reduction in $\alpha_{\rm IR}$ produces [Table~\ref{TT08} and Eq.\,\eqref{Efactor}]
\begin{equation}
\begin{array}{l|rrrr}
M\to M_<&\delta_T u & \delta_T d & g_T^{(0)} & g_T^{(1)}\\
\zeta_2 & 0.44 & -0.12 & 0.32 & 0.56
\end{array}\,,
\end{equation}
which expresses a 20\% decrease in $\delta_T u$.  As we signalled, the greatest impact of the cut in $\alpha_{\rm IR}$ and hence $M$ is a reduction in the size of the contributions from Diagrams~1 and 4: the former describes the tensor probe interacting with a dressed-quark whilst a scalar diquark is a spectator; and the latter involves a tensor probe exploring an axial-vector diquark with a dressed-quark bystander.

As remarked in the Introduction, the tensor charge is a defining intrinsic property of the proton and hence there is great interest in its reliable experimental and theoretical determination.  In Fig.\,\ref{TensorAll} we therefore compare our predictions with results from other analyses \cite{Hecht:2001ry,Bacchetta:2012ty,Cloet:2007em,Pasquini:2006iv,Wakamatsu:2007nc,%
alexandrou:2014,Gockeler:2005cj,Gamberg:2001qc,He:1994gz}.  Evidently, of all available computations, our contact-interaction predictions are in best agreement with the phenomenological estimates in Eq.\,\eqref{tensorud}.

Another interesting point is highlighted by a comparison between our predictions and the values obtained when the proton is considered to be a weakly-interacting collection of three massive valence-quarks described by an SU$(4)$-symmetric spin-flavour wave function \cite{He:1994gz}: $\delta_T^{{\rm SU}(4)} u =2 e_u$ and $\delta_T^{{\rm SU}(4)} d=e_d$ cf.\ our results, Eq.\,\eqref{errorResults}, $\delta_T u = 0.52 (2 e_u)$, $\delta_T d = 0.42(e_d)$.  The presence of diquark correlations in the proton amplitude significantly suppresses the magnitude of the tensor charge associated with each valence quark whilst simultaneously increasing the ratio $\delta_T d/\delta_T u$ by approximately 20\%.

%%...rescaling changed this outcome
%%In particular, consider the ratios:
%%\begin{equation}
%%\left.\frac{\delta_T d}{\delta_T u}\right|_{{\rm SU}(4)} = \frac{e_d}{2 e_u}
%%\quad {\rm cf}.\quad \left.\frac{\delta_T d}{\delta_T u}\right|_{{\rm Eq}.\,\eqref{tensorz2}} = \frac{e_d}{2 e_u}\,1.61\,.
%%\end{equation}
% 0.64 -0.17
% 1.33 -0.33
%%Thus, as we saw with the $\sigma$-term, the presence of $0^+$ and $1^+$ diquark correlations in the nucleon's Faddeev amplitude acts to dramatically increase the strength of the $d$-quark's contribution to the proton's tensor charge relative to that of the $u$-quark.

\section{Electric Dipole Moments}
\label{SecEDM}
In typical extensions of the Standard Model, quarks acquire an EDM \cite{Pospelov:2005pr,RamseyMusolf:2006vr}; i.e., an interaction with the photon that proceeds via a current of the form:
\begin{equation}
\tilde d_q \, q \gamma_5 \sigma_{\mu\nu} q\,,
\end{equation}
where $\tilde d_q$ is the quark's EDM and here we consider $q=u,d$.  The EDM of a proton containing quarks which interact in this way is defined as follows:
\begin{equation}
\langle P(p,\sigma) | \mathcal J_{\mu\nu}^\text{EDM} | P(p,\sigma)\rangle
= \tilde d_p\, \bar {\mathpzc u}(p,\sigma)\gamma_5\sigma_{\mu\nu} {\mathpzc u}(p,\sigma)\,,
\end{equation}
where
\begin{equation}
 \mathcal J_{\mu\nu}^\text{EDM}(x) = \tilde d_u\,\bar u(x)\gamma_5\sigma_{\mu\nu}u(x) + \tilde d_d\,\bar d(x)\gamma_5\sigma_{\mu\nu}d(x)\,.
\end{equation}

At this point it is useful to recall a simple Dirac-matrix identity:
\begin{equation}
\label{Epsilon}
\gamma_5\sigma_{\mu\nu} = \frac{1}{2} \varepsilon_{\mu\nu\alpha\beta} \sigma_{\alpha\beta}\,, \end{equation}
using which one can write
\begin{equation}
\mathcal J_{\mu\nu}^\text{EDM}
= \frac{1}{2} \varepsilon_{\mu\nu\alpha\beta}
\left[ \tilde d_u\,\bar u \sigma_{\alpha\beta}u  + \tilde d_d\,\bar d\sigma_{\alpha\beta}d\right]\,.
\end{equation}
It follows that
\begin{align}
&\langle P(p,\sigma) | \mathcal J_{\mu\nu}^\text{EDM} | P(p,\sigma)\rangle \nonumber\\
&= \frac{1}{2} \varepsilon_{\mu\nu\alpha\beta}
\left[
\tilde d_u \, \delta_T u\,
+ \tilde d_d \, \delta_T d\,
\right]\bar {\mathpzc u}(p,\sigma)\sigma_{\alpha\beta}{\mathpzc u}(p,\sigma) \\
&= \left[
\tilde d_u \, \delta_T u\,
+ \tilde d_d \, \delta_T d\,
\right]\bar {\mathpzc u}(p,\sigma)\gamma_5\sigma_{\mu\nu}{\mathpzc u}(p,\sigma) \,;
\end{align}
namely, the quark-EDM contribution to a proton's EDM is completely determined once the proton's tensor charges are known:
\begin{equation}
\tilde d_p = \tilde d_u \, \delta_T u\, + \tilde d_d \, \delta_T d\,.
\end{equation}
With emerging techniques, it is becoming possible to place competitive upper-limits on the proton's EDM using storage rings in which polarized particles are exposed to an electric field \cite{Pretz:2013us}.

An analogous result for the neutron is readily inferred.  In the limit of isospin symmetry,
\begin{align}
\langle N(p,\sigma)|\bar u\sigma_{\mu\nu}u|N(p,\sigma)\rangle &= \langle P(p,\sigma)|\bar d\sigma_{\mu\nu}d|P(p,\sigma)\rangle\,, \nonumber\\
\langle N(p,\sigma)|\bar d\sigma_{\mu\nu}d|N(p,\sigma)\rangle &= \langle P(p,\sigma)|\bar u\sigma_{\mu\nu}u|P(p,\sigma)\rangle\,;
\end{align}
and hence
\begin{equation}
\tilde d_n = \tilde d_u \, \delta_T d\, + \tilde d_d \, \delta_T u\,.
\end{equation}

Using the results in Eq.\,\eqref{errorResults}, we therefore have
\begin{equation}
\label{dndp}
\tilde d_n = -0.14\,\tilde d_u + 0.69\,\tilde d_d\,,\;
\tilde d_p = 0.69\,\tilde d_u - 0.14\,\tilde d_d\,.
\end{equation}

It is worth contrasting Eqs.\,\eqref{dndp} with the results one would obtain by assuming that the nucleon is merely a collection of three massive valence-quarks described by an SU$(4)$-symmetric spin-flavour wave function.  Then, by analogy with magnetic moment computations, a procedure also made valid by Eq.\,\eqref{Epsilon}:
\begin{equation}
\tilde d_n = -\frac{1}{3} \,\tilde d_u + \frac{4}{3} \,\tilde d_d\,,\;
\tilde d_p = \frac{4}{3} \,\tilde d_u - \frac{1}{3} \,\tilde d_d\,,
\end{equation}
%% ... once more, destroyed by rescaling ... Again, the presence of scalar and axial-vector diquark correlations within the nucleon enhances the effect of the singly-represented quark's EDM on that of the composite nucleon.
values which are roughly twice the size that we obtain.

The impact of our predictions for the scalar and tensor charges on BSM phenomenology may be elucidated, e.g., by following the analysis in Refs.\,\cite{Bhattacharya:2011qm,Dekens:2014jka}.

\section{Conclusion}
\label{SecConclusion}
We employed a confining, symmetry-preserving, Dyson-Schwinger equation treatment of a vector$\,\otimes\,$vector contact interaction in order to compute the dressed-quark-core contribution to the nucleon $\sigma$-term and tensor charges.  The latter enabled us to determine the effect of dressed-quark electric dipole moments (EDMs) on the neutron and proton EDMs.

A characteristic feature of DSE treatments of ground-state baryons is the predicted presence of strong scalar and axial-vector diquark correlations within these bound-states.  Indeed, in some respects the baryons can be viewed as weakly interacting dressed-quark$\,+\,$diquark composites.  The diquark correlations are active participants in all scattering events and therefore serve to modify the contribution to observables of the singly-represented valence-quark relative to that of the doubly-represented quark.

Regarding our analysis of the proton's $\sigma$-term, we estimate that with a realistic $d$-$u$ mass splitting, the singly-represented $d$-quark contributes 37\% more than the doubly-represented $u$-quark to that part of the proton mass which owes to explicit chiral symmetry breaking [Eqs.\,\eqref{split1}, \eqref{FLAGresult}].

Our predictions for the proton's tensor charges, $\delta_T u$, $\delta_T d$, are presented in Eq.\,\eqref{tensorz2}.  In this case, compared to results obtained in simple quark models, diquark correlations act to reduce the size of $\delta_T u$, $\delta_T d$ by a factor of two and increase the ratio $\delta_T d/\delta_T u$ by roughly 20\%.  Two additional observations are particularly significant.  First, the magnitude of $\delta_T u$ is a direct measure of the strength of DCSB in the Standard Model, diminishing rapidly with the difference between the scales of dynamical and explicit chiral symmetry breaking.  Second, $\delta_T d$ measures the strength of axial-vector diquark correlations in the proton, vanishing with $P_{1^+}/P_{0^+}$; i.e., the ratio of axial-vector- and scalar-diquark interaction probabilities, which is also a measure of DCSB.
%vanishing if such correlations are absent.

Our analysis of the Faddeev equation employed a simplifying truncation; viz., a variant of the so-called static approximation.  A natural next step is recalculation of the tensor charges after eliminating that truncation.  Subsequently or simultaneously, one might also employ the approaches of Refs.\,\cite{Eichmann:2013afa,Segovia:2014aza} in order to obtain DSE predictions with a more direct connection to QCD.

\begin{acknowledgments}
We thank Jian-ping Chen, Ian~Clo\"et, Haiyan~Gao, Michael Ramsey-Musolf, Jorge~Segovia, Ross~Young and Shu-sheng~Xu for insightful comments.
CDR acknowledges support of an \emph{International Fellow Award} from the Helmholtz Association.
Work otherwise supported by:
Austrian ``Fonds zur F\"orderung der Wissenschaftlichen Forschung'' (FWF) under contract no.~I689-N16;
U.S.\ Department of Energy, Office of Science, Office of Nuclear Physics, under contract nos.~DE-SC0011095 and DE-AC02-06CH11357;
and For\-schungs\-zentrum J\"ulich GmbH.
\end{acknowledgments}

\appendix
\setcounter{figure}{0}
\renewcommand{\thefigure}{\Alph{section}.\arabic{figure}}
\renewcommand{\thetable}{\Alph{section}.\arabic{table}}

\section{Contact interaction}
\label{sec:contact}
Our treatment of the contact interaction begins with the gap equation
\begin{eqnarray}
\nonumber \lefteqn{S(p)^{-1}= i\gamma\cdot p + m}\\
&&+ \!\! \int \! \frac{d^4q}{(2\pi)^4} g^2 D_{\mu\nu}(p-q) \frac{\lambda^a}{2}\gamma_\mu S(q) \frac{\lambda^a}{2}\Gamma_\nu(q,p) ,\;
\label{gendse}
\end{eqnarray}
wherein $m$ is the Lagrangian current-quark mass, $D_{\mu\nu}$ is the vector-boson propagator and $\Gamma_\nu$ is the quark--vector-boson vertex.  We work with the choice
\begin{equation}
\label{njlgluon}
%g^2 D_{\mu \nu}(p-q) = \delta_{\mu \nu} \frac{1}{m_G^2}\,,
g^2 D_{\mu \nu}(p-q) = \delta_{\mu \nu} \frac{4 \pi \alpha_{\rm IR}}{m_G^2}\,,
\end{equation}
%where $m_G$ is a gluon mass-scale,
where $m_G=0.8\,$GeV is a gluon mass-scale typical of the one-loop renormalisation-group-improved interaction introduced in Ref.\,\cite{Qin:2011dd}, and the fitted parameter $\alpha_{\rm IR}/\pi = 0.93$ is commensurate with contemporary estimates of the zero-momentum value of a running-coupling in QCD \cite{Aguilar:2010gm,Boucaud:2010gr}.  Equation~\eqref{njlgluon} is embedded in a rainbow-ladder (RL) truncation of the DSEs, which is the leading-order in the most widely used, global-symmetry-preserving truncation scheme \cite{Munczek:1994zz,Bender:1996bb}.  This means
\begin{equation}
\label{RLvertex}
\Gamma_{\nu}(p,q) =\gamma_{\nu}
\end{equation}
in Eq.\,(\ref{gendse}) and in the subsequent construction of the Bethe-Salpeter kernels.

One may view the interaction in Eq.\,(\ref{njlgluon}) as being inspired by models of the Nambu--Jona-Lasinio type \cite{Nambu:1961tp}.  However, our treatment is atypical.  Moreover, as noted in the Introduction, one normally finds Eqs.\,\eqref{njlgluon}, (\ref{RLvertex}) produce results for low-momentum-transfer observables that are practically indistinguishable from those produced by more sophisticated interactions
\cite{GutierrezGuerrero:2010md,Roberts:2010rn,Roberts:2011cf,Roberts:2011wy,Wilson:2011aa,%
Chen:2012qr,Chen:2012txa,Pitschmann:2012by,Wang:2013wk,Segovia:2013rca,Segovia:2013uga}.
Using Eqs.\,(\ref{njlgluon}), (\ref{RLvertex}), the gap equation becomes
\begin{equation}
 S^{-1}(p) =  i \gamma \cdot p + m +  \frac{16\pi}{3}\frac{\alpha_{\rm IR}}{m_G^2} \int\!\frac{d^4 q}{(2\pi)^4} \,
\gamma_{\mu} \, S(q) \, \gamma_{\mu}\,,   \label{gap-1}
\end{equation}
an equation in which the integral possesses a quadratic divergence.  When the divergence is regularised in a Poincar\'e covariant manner, the solution is
\begin{equation}
\label{genS}
S(p)^{-1} = i \gamma\cdot p + M\,,
\end{equation}
where $M$ is momentum-independent and determined by
\begin{equation}
M = m + M\frac{4\alpha_{\rm IR}}{3\pi m_G^2} \int_0^\infty \!ds \, s\, \frac{1}{s+M^2}\,.
\end{equation}

We define Eq.\,\eqref{gap-1} by writing \cite{Ebert:1996vx}
\begin{eqnarray}
\nonumber
\frac{1}{s+M^2} & = & \int_0^\infty d\tau\,{\rm e}^{-\tau (s+M^2)} \\
& \rightarrow & \int_{\tau_{\rm uv}^2}^{\tau_{\rm ir}^2} d\tau\,{\rm e}^{-\tau (s+M^2)}
\label{RegC}\\
& & =
\frac{{\rm e}^{- (s+M^2)\tau_{\rm uv}^2}-e^{-(s+M^2) \tau_{\rm ir}^2}}{s+M^2} \,, \label{ExplicitRS}
\end{eqnarray}
where $\tau_{\rm ir,uv}$ are, respectively, infrared and ultraviolet regulators.  It is apparent from Eq.\,(\ref{ExplicitRS}) that a finite value of $\tau_{\rm ir}=:1/\Lambda_{\rm ir}$ implements confinement by ensuring the absence of quark production thresholds \cite{Krein:1990sf}.  Since Eq.\,(\ref{njlgluon}) does not define a renormalisable theory, then $\Lambda_{\rm uv}:=1/\tau_{\rm uv}$ cannot be removed but instead plays a dynamical role, setting the scale of all dimensioned quantities.  Using Eq.\,\eqref{RegC}, the gap equation becomes
\begin{equation}
%M = m +  \frac{M}{3\pi^2 m_G^2} \,{\cal C}(M^2;\tau_{\rm ir},\tau_{\rm uv})\,,
%M = m +  \frac{M}{3\pi^2 m_G^2} \,{\cal C}^{\rm iu}(M^2)\,,
M = m + M\frac{4\alpha_{\rm IR}}{3\pi m_G^2}\,\,{\cal C}^{\rm iu}(M^2)\,,
\label{gapactual}
\end{equation}
where,
\begin{align}
  \mathcal C^\text{iu}(\omega)
  %&= \omega \bar{\mathcal C}^\text{iu}(\omega)
  &=\omega\,[\Gamma(-1,\omega \tau_{\rm uv}^2) - \Gamma(-1,\omega\tau_{\rm ir}^2)]\,,
\end{align}
with $\Gamma(\alpha,y)$ being the incomplete gamma-function.

At this point we also list expressions for the other regularised integrals that we employ herein:
{\allowdisplaybreaks
\begin{align}
  \mathcal C_{n}^\text{iu}(\omega)&=(-1)^n\frac{\omega^n}{n!}\frac{d^n}{d\omega^n}\,\mathcal C^\text{iu}(\omega)\,, \\
  \mathcal D^{\rm iu}(\omega) &= \int_R ds\,\frac{s^2}{s + M^2} \nonumber \\
  & = 2\omega^2\,[\Gamma(-2,\omega\tau_{\rm uv}^2) - \Gamma(-2,\omega\tau_{\rm ir}^2)]\,,  \\
  \mathcal E^{\rm iu}(\omega) &= \int_R ds\,\frac{s^3}{s + M^2} \nonumber \\
  & = 6\omega^3\,[\Gamma(-3,\omega\tau_{\rm uv}^2) - \Gamma(-3,\omega\tau_{\rm ir}^2)]\,,\\
\label{DFG1}
  \check{\mathcal G}_1^{\rm iu}(\omega) &= \int_R ds\,\frac{s}{\left(s + \omega\right)^3} = \frac{1}{2}\frac{d^2}{d\omega^2}\,\mathcal C^\text{iu}(\omega)\,, \\
\label{DFG2}
  \check{\mathcal G}_2^{\rm iu}(\omega) &= \int_R ds\,\frac{s^2}{\left(s + \omega\right)^3} \nonumber\\
  & = \bar{\mathcal C}_1^\text{iu}(\omega) - \frac{\omega}{2}\frac{d^2}{d\omega^2}\,\mathcal C^\text{iu}(\omega)\,, \\
  \check{\mathcal G}_3^{\rm iu}(\omega) &= \int_R ds\,\frac{s^3}{\left(s + \omega\right)^3} \nonumber \\
    &= \mathcal C^\text{iu}(\omega) - 2\,\mathcal C_1^\text{iu}(\omega) + \mathcal C_2^\text{iu}(\omega)\,, \\
  \check{\mathcal G}_4^{\rm iu}(\omega) &= \int_R ds\,\frac{s^4}{\left(s + \omega\right)^3} = \mathcal D^{\rm iu}(\omega) \nonumber \\
  & \quad - 2\omega\,\mathcal C^\text{iu}(\omega) + 3\omega\,\mathcal C_1^\text{iu}(\omega) - \omega\,\mathcal C_2^\text{iu}(\omega)\,, \\
  \check{\mathcal G}_5^{\rm iu}(\omega) &= \int_R ds\,\frac{s^5}{\left(s + \omega\right)^3} = \mathcal E^{\rm iu}(\omega) - 2\omega\,\mathcal D^{\rm iu}(\omega)  \nonumber \\
    & \quad + 3\omega^2\,\mathcal C^\text{iu}(\omega) - 4\omega^2\,\mathcal C_1^\text{iu}(\omega) + \omega^2\,\mathcal C_2^\text{iu}(\omega)\,,
\end{align}}
where $\{{\mathcal G}_i = \check{\mathcal G}_i/(16\pi^2)$, $i=1,\ldots,5\}$.

The parameters that specify our treatment of the contact interaction were determined in a study of $\pi$- and $\rho$-meson properties \cite{Roberts:2011wy}; viz., $\alpha_{\rm IR}/\pi=0.93$ and (in GeV)
\begin{equation}
\label{parametervalues}
m=0.007\,,\;
\Lambda_{\rm ir} = 0.240\,\;
\Lambda_{\rm uv}=0.905\,,
\end{equation}
using which, Eq.\,\eqref{gapactual} yields
\begin{equation}
\label{dressedMvalue}
M = 0.368\,{\rm GeV}.
\end{equation}

With the aim of exploring the impact of DCSB on our results, herein we also consider results obtained with $\alpha_{\rm IR}/\pi=0.74$, in which case
\begin{equation}
\label{dressedMvalue08}
M \to M_{<} = 0.246\,{\rm GeV}.
\end{equation}

\section{Faddeev Equation}
\label{sec:Faddeev}
We describe the dressed-quark-cores of the nucleon via solutions of a Poincar\'e-covariant Faddeev equation \cite{Cahill:1988dx}.  The equation is derived following upon the observation that an interaction which describes mesons also generates quark-quark (diquark) correlations in the colour-$\bar 3$ channel \cite{Cahill:1987qr}.  The fidelity of the diquark approximation to the quark-quark scattering kernel has been  verified \cite{Eichmann:2011vu}.

In RL truncation, the colour-antitriplet diquark correlations are described by an homogeneous Bethe-Salpeter equation that is readily inferred from the analogous meson equation; viz., following Ref.\,\cite{Cahill:1987qr} and expressing the diquark amplitude as
\begin{equation}
\Gamma^c_{qq}(k;P) = \Gamma_{qq}(k;P) C^\dagger H^{c},
\end{equation}
with
\begin{equation}
\label{Hmatrices}
\{H^1=i\lambda^7,H^2=-i\lambda^5,H^3=i\lambda^2\}\,, \epsilon_{c_1 c_2 c_3}= (H^{c_3})_{c_1 c_2}\,,
\end{equation}
%[See Eqs.\,(\ref{Gamma0p}), (\ref{Gamma1p}).]
where $\{\lambda^{2,5,7}\}$ are Gell-Mann matrices, then
\begin{equation}
\Gamma_{qq}(k;P) = -\frac{8 \pi }{3}\frac{\alpha_{\rm IR}}{m_G^2} \int \! \frac{d^4q}{(2\pi)^4}\, \gamma_\mu \chi_{qq}(q;P)\gamma_\mu \,,
\label{genbseqq}
\end{equation}
where $\chi_{qq}(q;P)=S(q)\Gamma_{qq}(P)S(q-P)$ and $\Gamma_{qq}$ is the diquark Bethe-Salpeter amplitude, which is independent of the relative momentum when using a contact interaction \cite{Roberts:2011wy}.%It follows that one may obtain the mass and amplitude for a diquark with spin-parity $J^P$ from the equation for a $J^{-P}$-meson in which the only change is a halving of the interaction strength.  The flipping of the sign in parity occurs because fermions and antifermions have opposite parity.

Scalar and axial-vector quark-quark correlations are dominant in studies of the nucleon:
\begin{eqnarray}
\label{scqqbsa}
\Gamma^{0^+}_{qq}(P) &= &  i \gamma_5 E_{qq 0}(P) + \frac{1}{M} \gamma_5 \gamma\cdot P F_{qq 0}(P) \,,\rule{2em}{0ex}\\
i\Gamma_{qq\, \mu}^{1^+}(P) & = & i\gamma^T_\mu E_{qq 1}(P),\label{avqqbsa}
\end{eqnarray}
where $P_\mu \gamma^T_\mu = 0$.  These amplitudes are canonically normalised:
\begin{equation}
P_\mu = 2 {\rm tr}\!\!\int\frac{d^4q}{(2\pi)^4} \Gamma_{qq}^{0^+}(-P) \frac{\partial}{\partial P_\mu} S(q+P) \Gamma_{qq}^{0^+}(P) S(q);
\end{equation}
and
\begin{equation}
P_\mu = \frac{2}{3} {\rm tr}\!\!\int\frac{d^4q}{(2\pi)^4} \Gamma_{qq\,\alpha}^{1^+}(-P) \frac{\partial}{\partial P_\mu} S(q+P) \Gamma_{qq\,\alpha}^{1^+}(P) S(q).
\label{avqqNorm}
\end{equation}

A $J=\frac{1}{2}$ baryon is represented by a Faddeev amplitude
\begin{equation}
\label{PsiNucleon}
\Psi = \Psi_1 + \Psi_2 + \Psi_3  \,,
\end{equation}
where the subscript identifies the bystander quark and, e.g., $\Psi_{1,2}$ are obtained from $\Psi_3$ by a cyclic permutation of all the quark labels.  We employ a simple but realistic representation of $\Psi$.  The spin- and isospin-$\frac{1}{2}$ nucleon is a sum of scalar and axial-vector diquark correlations:
\begin{equation}
\label{Psi} \Psi_3(p_i,\alpha_i,\tau_i) = {\cal N}_3^{0^+} + {\cal N}_3^{1^+},
\end{equation}
with $(p_i,\alpha_i,\tau_i)$ the momentum, spin and isospin labels of the
quarks constituting the bound state, and $P=p_1+p_2+p_3$ the system's total momentum.

The scalar diquark piece in Eq.\,(\ref{Psi}) is
\begin{align}
\nonumber
{\cal N}_3^{0^+}(p_i,\alpha_i,\tau_i) &= [\Gamma^{0^+}(\frac{1}{2}p_{[12]};K)]_{\alpha_1
\alpha_2}^{\tau_1 \tau_2}\\
& \quad \times \Delta^{0^+}(K) \,[{\cal S}(\ell;P) {\mathpzc u}(P)]_{\alpha_3}^{\tau_3},
\label{calS}
\end{align}
where: the spinor satisfies Eq.\,(\ref{DiracEqn}), with $M$ the mass obtained by solving the Faddeev equation, and it is also a spinor in isospin space with $\varphi_+= {\rm col}(1,0)$ for the charge-one state and $\varphi_-= {\rm col}(0,1)$ for the neutral state; $K= p_1+p_2=: p_{\{12\}}$, $p_{[12]}= p_1 - p_2$, $\ell := (-p_{\{12\}} + 2 p_3)/3$;
\begin{equation}
\label{scalarqqprop}
\Delta^{0^+}(K) = \frac{1}{K^2+m_{qq_{0^+}}^2}
\end{equation}
is a propagator for the scalar diquark formed from quarks $1$ and $2$, with $m_{qq_{0^+}}$ the mass-scale associated with this correlation, and $\Gamma^{0^+}\!$ is the canonically-normalised Bethe-Salpeter amplitude described above; and ${\cal S}$, a $4\times 4$ Dirac matrix, describes the relative quark-diquark momentum correlation.

The axial-vector component in Eq.\,(\ref{Psi}) is
\begin{align}
\nonumber
{\cal N}^{1^+}(p_i,\alpha_i,\tau_i) & =  [{\tt t}^i\,\Gamma_\mu^{1^+}(\frac{1}{2}p_{[12]};K)]_{\alpha_1
\alpha_2}^{\tau_1 \tau_2}\\
& \quad\times \Delta_{\mu\nu}^{1^+}(K)\,
[{\cal A}^{i}_\nu(\ell;P) {\mathpzc u}(P)]_{\alpha_3}^{\tau_3}\,,
\label{calA}
\end{align}
where the symmetric isospin-triplet matrices are
\begin{equation}
{\tt t}^+ = \frac{1}{\surd 2}(\tau^0+\tau^3) \,,\;
{\tt t}^0 = \tau^1\,,\;
{\tt t}^- = \frac{1}{\surd 2}(\tau^0-\tau^3)\,,
\end{equation}
and the other elements in Eq.\,(\ref{calA}) are straightforward generalisations of those in Eq.\,(\ref{calS}) with, e.g.,
\begin{equation}
\label{avqqprop}
\Delta_{\mu\nu}^{1^+}(K) = \frac{1}{K^2+m_{qq_{1^+}}^2} \, \left(\delta_{\mu\nu} + \frac{K_\mu K_\nu}{m_{qq_{1^+}}^2}\right) \,.
\end{equation}

One can now write the Faddeev equation for $\Psi_3$:
\begin{eqnarray}
\nonumber
\lefteqn{
 \left[ \begin{array}{r}
{\cal S}(k;P)\, {\mathpzc u}(P)\\
{\cal A}^i_\mu(k;P)\, {\mathpzc u}(P)
\end{array}\right]}\\
& =&  -\,4\,\int\frac{d^4\ell}{(2\pi)^4}\,{\cal M}(k,\ell;P)
\left[
\begin{array}{r}
{\cal S}(\ell;P)\, {\mathpzc u}(P)\\
{\cal A}^j_\nu(\ell;P)\, {\mathpzc u}(P)
\end{array}\right] .\rule{1em}{0ex}
\label{FEone}
\end{eqnarray}
The kernel in Eq.\,(\ref{FEone}) is
\begin{equation}
\label{calM} {\cal M}(k,\ell;P) = \left[\begin{array}{cc}
{\cal M}_{00} & ({\cal M}_{01})^j_\nu \\
({\cal M}_{10})^i_\mu & ({\cal M}_{11})^{ij}_{\mu\nu}\rule{0mm}{3ex}
\end{array}
\right] ,
\end{equation}
with
\begin{eqnarray}
\nonumber
 {\cal M}_{00} &=& \Gamma^{0^+}\!(k_q-\ell_{qq}/2;\ell_{qq})\,
S^{\rm T}(\ell_{qq}-k_q) \\
&& \times \,\bar\Gamma^{0^+}\!(\ell_q-k_{qq}/2;-k_{qq})\,
S(\ell_q)\,\Delta^{0^+}(\ell_{qq}) \,, \rule{2em}{0ex}
\end{eqnarray}
where: $\ell_q=\ell$, $k_q=k$, $\ell_{qq}=-\ell+ P$,
$k_{qq}=-k+P$, the superscript ``T'' denotes matrix transpose, $\bar\Gamma$ is defined in Eq.\,\eqref{chargec}; and
\begin{eqnarray}
\nonumber
({\cal M}_{01})^j_\nu &=& {\tt t}^j \,
\Gamma_\mu^{1^+}\!(k_q-\ell_{qq}/2;\ell_{qq}) S^{\rm T}(\ell_{qq}-k_q)\,\\
&& \rule{-1.5em}{0ex} \times \bar\Gamma^{0^+}\!(\ell_q-k_{qq}/2;-k_{qq})\,
S(\ell_q)\,\Delta^{1^+}_{\mu\nu}(\ell_{qq}) , \rule{2.2em}{0ex} \label{calM01} \\
\nonumber
({\cal M}_{10})^i_\mu &=& \Gamma^{0^+}\!(k_q-\ell_{qq}/2;\ell_{qq})\,
S^{\rm T}(\ell_{qq}-k_q)\,{\tt t}^i\, \\
&& \rule{-1.5em}{0ex}\times \bar\Gamma_\mu^{1^+}\!(\ell_q-k_{qq}/2;-k_{qq})\,
S(\ell_q)\,\Delta^{0^+}(\ell_{qq}) , \rule{2.2em}{0ex}\\
\nonumber
({\cal M}_{11})^{ij}_{\mu\nu} &=& {\tt t}^j\,
\Gamma_\rho^{1^+}\!(k_q-\ell_{qq}/2;\ell_{qq})\, S^{\rm T}(\ell_{qq}-k_q)\,{\tt t}^i\,\\
&& \rule{-1.5em}{0ex}\times  \bar\Gamma^{1^+}_\mu\!(\ell_q-k_{qq}/2;-k_{qq})\,
S(\ell_q)\,\Delta^{1^+}_{\rho\nu}(\ell_{qq}) . \rule{2.2em}{0ex}\label{calM11}
\end{eqnarray}

The dressed-quark propagator is described in Sec.\,\ref{sec:contact} and the diquark propagators are given in Eqs.\,(\ref{scalarqqprop}), (\ref{avqqprop}), so the Faddeev equation is complete once the diquark Bethe-Salpeter amplitudes are computed from Eqs.\,\eqref{genbseqq} -- \eqref{avqqNorm}.  However, we follow Ref.\,\cite{Roberts:2011cf} and employ a simplification of the kernel; viz., in the Faddeev equation, the quark exchanged between the diquarks is represented as
\begin{equation}
S^{\rm T}(k) \to \frac{g_N^2}{M}\,,
\label{staticexchange}
\end{equation}
where $g_N=1.18$.  This is a variant of the so-called ``static approximation,'' which itself was introduced in Ref.\,\cite{Buck:1992wz} and has subsequently been used in studying a range of nucleon properties \cite{Bentz:2007zs}.  In combination with diquark correlations generated by Eq.\,(\ref{njlgluon}), whose Bethe-Salpeter amplitudes are momentum-independent, Eq.\,(\ref{staticexchange}) generates Faddeev equation kernels which themselves are momentum-independent.  The dramatic simplifications which this produces are the merit of Eq.\,(\ref{staticexchange}).  Nevertheless, we are currently exploring the veracity of this truncation.

The general forms of the matrices ${\cal S}(\ell;P)$ and ${\cal A}^i_\nu(\ell;P)$, which describe the momentum-space correlation between the quark and diquark in the nucleon, are described in Refs.\,\cite{Oettel:1998bk,Cloet:2007pi}.  However, with the interaction described in Sec.\,\ref{sec:contact} augmented by Eq.\,(\ref{staticexchange}), they simplify greatly; viz.,
\begin{subequations}
\label{FaddeevAmp}
\begin{eqnarray}
{\cal S}(P) &=& s(P) \,{\mathbb 1}\,,\\
i{\cal A}^j_\mu(P) &=& a_1^j(P) \gamma_\mu\gamma_5 + i a_2^j(P) \gamma_5 \hat P_\mu \,,j=+,0\,,\rule{2em}{0ex}
\end{eqnarray}
\end{subequations}
with the scalars $s$, $a_{1,2}^i$ independent of the relative quark-diquark momentum and $\hat P^2=-1$.

The mass of the ground-state nucleon is then determined by a $5\times 5$ matrix Faddeev equation; viz., $\Psi = K \Psi$, with the eigenvector defined via
\begin{equation}
\Psi(P)^{\rm T} = \left[
s(P) \;
a_1^+(P)\;
a_1^0(P)\;
a_2^+(P)\;
a_2^0(P)\right],
\end{equation}
and the kernel $({\mathpzc k}_\pm = \pm \surd 2)$
%\begin{widetext}
\begin{align}
& K(P) =\nonumber\\
& \left[ \begin{array}{ccccc}
K^{00}_{ss} & {\mathpzc k}_- \, K^{01}_{sa_1} & K^{01}_{sa_1} & {\mathpzc k}_-\, K^{01}_{sa_2} & K^{01}_{sa_2}\\[0.7ex]
{\mathpzc k}_-\, K^{10}_{a_1 s} & 0 & {\mathpzc k}_+\, K^{11}_{a_1 a_1} & 0 & {\mathpzc k}_+\, K^{11}_{a_1 a_2}\\[0.7ex]
K^{10}_{a_1 s} & {\mathpzc k}_+ \, K^{11}_{a_1 a_1} & K^{11}_{a_1 a_1} & {\mathpzc k}_+\,K^{11}_{a_1 a_2} & K^{11}_{a_1 a_2} \\[0.7ex]
{\mathpzc k}_-\, K^{10}_{a_2 s} & 0 & {\mathpzc k}_+\, K^{11}_{a_2 a_1} & 0 & {\mathpzc k}_+\, K^{11}_{a_2 a_2} \\[0.7ex]
K^{10}_{a_2 s} & {\mathpzc k}_+\, K^{11}_{a_2 a_1} & K^{11}_{a_2 a_1} & {\mathpzc k}_+\, K^{11}_{a_2 a_2} & K^{11}_{a_2 a_2}
\end{array}
\right],
\end{align}
%\end{widetext}
whose entries are given explicitly in Eqs.\,(B20), (B21) of Ref.\,\cite{Wilson:2011aa}. Given the structure of the kernel, the eigenvectors exhibit the pattern:
\begin{equation}
\label{aia0}
a_i^+ = -\sqrt{ 2} a_i^0,\; i=1,2.
\end{equation}

Using the parameters and results described in and connection with Eqs.\,\eqref{parametervalues}, \eqref{dressedMvalue}, the diquark Bethe-Salpeter equations produce the following diquark masses (in GeV)
\begin{equation}
m_{qq {0^+}} = 0.78\,,\; m_{qq {1^+}} = 1.06\,,
\end{equation}
and canonically normalised amplitudes:
\begin{equation}
\label{qqCABSA}
%false ... typo in Wilson E_{qq {0^+}} = 4.351 \,,\;
%false ... typo in Wilson ... F_{qq {0^+}} = 0.498 \,,\;
E_{qq {0^+}} = 2.742 \,,\;
F_{qq {0^+}} = 0.314 \,,\;
E_{qq {1^+}} = 1.302 \,.
\end{equation}
With this input to the Faddeev equation, one obtains \cite{Roberts:2011cf,Wilson:2011aa,Chen:2012qr} $m_N=1.14\,$GeV and the following unit-normalised eigenvector\footnote{$E_{qq {0^+}}$, $F_{qq {0^+}}$ listed in Table~I(A) of Ref.\,\cite{Wilson:2011aa} are incorrect.  The values listed in Eq.\,\eqref{qqCABSA} were actually used therein.}
\begin{equation}
\label{NucleonEigenVector}
\begin{array}{ccccc}
s(P) & a_1^+(P) & a_1^0(P) & a_2^+(P) & a_2^0(P)\\
0.88 & -0.38 & 0.27 & -0.065  & 0.046
\end{array}\,.
\end{equation}
As explained elsewhere \cite{Roberts:2011cf,Wilson:2011aa,Chen:2012qr}, the mass is greater than that determined empirically because our Faddeev equation kernel omits resonant contributions; i.e., does not contain effects that may phenomenologically be associated with a meson cloud.  It is for this reason that our Faddeev equation describes the nucleon's dressed-quark core.  Notably, meson cloud effects typically work to reduce a hadron's mass \cite{Hecht:2002ej}.

%%% (2) ... reduced alpha
%%%  M_Q         =  0.245613
%%%  M_pi        =  0.149692
%%%  M_rho       =  0.862438
%%%  M_0+        =  0.696649
%%%  M_1+        =  0.981385
%%%  M_1+ -M_0+  =  0.284736
%%%  Es1C        =  2.16542 ... canonically normalised scalar-d amplitude "E"
%%%  Fs1C        =  0.138522 ... canonically normalised scalar-d amplitude "F"
%%%  EavdC        =  1.09272 ... ... canonically normalised axial-vector-diquark
%%%  M_Delta     =  1.26819
%%%  M_N         =  1.02175
%%%  M_Delta-M_N =  0.246436

%%% Nucleon Faddeev amplitude
%%% 0.878486 ; -0.384547 ; 0.271916 ;  -0.0655619 ; 0.0463593

Using the reduced coupling value described in connection with Eq.\,\eqref{dressedMvalue08}, the diquark Bethe-Salpeter equations produce the following diquark masses (in GeV)
\begin{equation}
m_{qq{0^+}} = 0.70\,,\; m_{qq{1^+}} = 0.98\,,
\end{equation}
and canonically normalised amplitudes:
\begin{equation}
E_{qq {0^+}} = 2.165 \,,\;
F_{qq {0^+}} = 0.139 \,,\;
E_{qq {1^+}} = 1.093 \,.
\end{equation}
With this input to the Faddeev equation, one obtains $m_N=1.02\,$GeV and the following unit-normalised eigenvector
\begin{equation}
\label{NucleonEigenVector08}
\begin{array}{ccccc}
s(P) & a_1^+(P) & a_1^0(P) & a_2^+(P) & a_2^0(P)\\
0.88 & -0.38 & 0.27 & -0.065  & 0.046
\end{array}\,.
\end{equation}
Plainly, a 20\% cut in the infrared value of the coupling diminishes the strength of DCSB by 33\%.  This feeds into reductions of the diquark Bethe-Salpeter amplitudes and a 10\% cut in the nucleon mass.  On the other hand, the nucleon's Faddeev amplitude, which describes its internal structure, is almost unchanged.  The same pattern is seen in studies of the temperature dependence of nucleon properties \cite{Wang:2013wk}.
%These results are the strength of DCSB is markedly reduced decreases the nucleon mass changes but the internal structure is almost unaltered.  I have seen this before; viz., the same thing happens with increasing temperature ... <http://arxiv.org/pdf/1301.6762.pdf>, Figs. 9 & 11.

\section{Interaction Currents}
\label{ICurrents}
In order to translate the diagrams drawn in this Appendix into formulae, it is helpful to bear the following points in mind.
\smallskip

\hspace*{-\parindent}(1) In front of a closed fermion trace; i.e., a vertex, one should, as usual, include a factor of $(-1)$.
%$ \\(necessary to get positive $\sigma$-term)
\smallskip

\hspace*{-\parindent}(2a) States entering a diagram are described by the amplitudes
\begin{subequations}
\begin{align}
\Gamma_{qq}^{0^+}(P) &= \gamma_5\,(i E_{qq 0^+} + \frac{1}{M}\,\gamma\cdot P\,F_{qq 0^+}) \,,\\
\Gamma_{qq \mu}^{1^+}(P) & = i E_{qq 1^+}\gamma_\mu^T \,,\\
\mathcal S(P)\, &= s\,{\mathbb 1}\,, \\
\mathcal A_\mu^j (P) &= a_1^j\gamma_\mu\gamma_5 + i a_2^j\gamma_5\hat P_\mu \,.
\end{align}
\end{subequations}
(N.B.\ In this Appendix we have absorbed the ``$i$'' of Eqs.\,\eqref{avqqbsa},  \eqref{FaddeevAmp} into the labels $\Gamma_{qq \mu}^{1^+}(P)$ and $\mathcal A_\mu^j$.)

\smallskip

\hspace*{-\parindent}(2b) States leaving a diagram are described by the amplitudes
{\allowdisplaybreaks
\begin{subequations}
\begin{align}
\Gamma_{qq}^{0^+}(-P) &= \gamma_5\,(i E_{qq 0^+} - \frac{1}{M}\,\gamma\cdot P\,F_{qq 0^+}) \,,\\
\Gamma_{qq \mu}^{1^+}(-P) & = i E_{qq 1^+}\gamma_\mu^T \,,\\
\mathcal S(-P)\, &= s\,{\mathbb 1} =: \bar{\mathcal S}\,, \\
\mathcal A_\mu^j (-P) &= a_1^j\gamma_5\gamma_\mu + i a_2^j\gamma_5\hat P_\mu \,.
\end{align}
\end{subequations}}
\smallskip

In these equations,
\begin{equation}
  \gamma_\mu^T = \gamma_\nu\mathbb P_{\mu\nu}(P)\,, \quad
  \mathbb P_{\mu\nu}(P) =\delta_{\mu\nu} + \frac{P_\mu P_\nu}{m_{qq1^+}^2}\,.
\end{equation}
\smallskip

\hspace*{-\parindent}(3) In the traces arising from a closed fermion loop, we have: $\bar e_j \bar N$ for charge form factors, where $\bar e_0 = \frac{1}{3}e$, $\bar e_+ = \frac{4}{3}e$, where $e$ is the positron charge; and $2\bar N$ for scalar and tensor form factors.  Note that $\bar N=2$ for diquark initial and final states.

\subsection{Electromagnetic Current}
\label{emApp}
In computing the charge form factor of any hadron, one must employ the dressed-quark-photon vertex \cite{Roberts:1994hh,Qin:2013mta}.  That vertex may be obtained by solving an inhomogeneous Bethe-Salpeter equation whose unrenormalised form is determined by the inhomogeneous term $\gamma_\mu$.  The complete solution for the contact-interaction's vector vertex in RL truncation can be found in Refs.\,\cite{Roberts:2010rn,Chen:2012txa}; but that result is not necessary herein because we only require the result at $Q^2=0$, which is fixed by the Ward identity.  With the contact interaction, that means
\begin{equation}
{\mathcal V}_\mu^q(Q) \stackrel{Q^2=0}{=} e_q \gamma_\mu\,,
\end{equation}
where $e_q$ is the quark's electric charge.

The $Q^2=0$ value of the elastic electromagnetic proton current determines the canonical normalisation of the nucleon's Faddeev amplitude \cite{Oettel:1999gc}.  Given the Faddeev equation in Fig.\,\ref{figFaddeev}, the complete result is obtained by summing the six one-loop diagrams that we now describe.  There would be more diagrams if the interaction were momentum dependent \cite{Oettel:1999gc}.

\subsubsection{Diagram 1 -- em}
\begin{figure}[t]
  \centering
  \vspace{1mm}
  \includegraphics[width=0.5\linewidth]{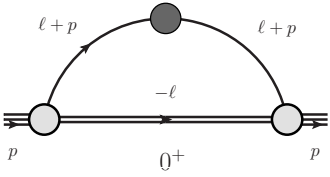}
  \caption{Diagram 1: The probe interacts with a quark within the proton and the $0^+$ diquark is a bystander.}
  \label{DiC1}
\end{figure}

The first contribution is depicted in Fig.\,\ref{DiC1}, which translates into the following expression
{\allowdisplaybreaks
\begin{eqnarray}
\lefteqn{e\,Q_{p,1}\,\Lambda^+(p)\gamma_\mu\Lambda^+(p) = \mathcal N\,\Lambda^+(p)\,\bar{\mathcal S}\,\int\frac{d^4\ell}{(2\pi)^4}\,}\nonumber\\
&& \times S(\ell + p)e_u\gamma_\mu S(\ell + p)\Delta^{0^+}(-\ell)\,\mathcal S\,\Lambda^+(p) \\
%
%  &= \mathcal N\,\Lambda^+(p)\,s^2\int\frac{d^4\ell}{(2\pi)^4}\frac{i\gamma\cdot(\ell + p) - M}{(\ell + p)^2 + M^2}\,e_u\gamma_\mu\,\frac{i\gamma\cdot(\ell + p) - M}{(\ell + p)^2 + M^2}\frac{1}{\ell^2 + m_{qq_0}^2}\,\Lambda^+(p) \nonumber\\
%
&= & 2\,\mathcal N\,\Lambda^+(p)\,s^2\int_0^1 dx\,(1 - x)\int\frac{d^4\ell}{(2\pi)^4} \nonumber \\
&& \frac{\{i\gamma\cdot(\ell + x p) - M\}\,e_u\gamma_\mu\,\{i\gamma\cdot(\ell + x p) - M\}}{[\ell^2 - x(1 - x)m_N^2 + (1 - x)M^2 + xm_{qq_0}^2]^3}\,\Lambda^+(p)\,,
\nonumber \\
\end{eqnarray}}
\hspace*{-0.5\parindent}
where here and hereafter we (often) suppress the parity-$+$ superscript on the diquark label, $\mathcal S$ is the scalar-diquark piece of the Faddeev amplitude and ${\cal N}$ is the (as yet undetermined) canonical normalisation constant for the Faddeev amplitude that ensures that the proton charge is unity; i.e., $Q_p=1$.

Applying the projection operator
\begin{align}
\label{eqProjectorCharge}
  \mathcal P_\mu = \frac{1}{2}\,\gamma_\mu\,,
\end{align}
and performing the trace, one obtains
\begin{eqnarray}
\lefteqn{e\,Q_{p,1} = e_u\,\mathcal N\,s^2\int_0^1 dx\,(1 - x)\int\frac{d^4\ell}{(2\pi)^4}}\nonumber \\
&& \frac{\ell^2 + 2(M + xm_N)^2}{[\ell^2 - x(1 - x)m_N^2 + (1 - x)M^2 + xm_{qq_0}^2]^3} \\
&\to& e_u\,\mathcal N\,s^2\int_0^1dx\,(1 - x)\bigg\{\mathcal G_2^{\rm iu}\Big(x(x - 1)m_N^2\nonumber \\
&& + (1 - x)M^2 + xm_{qq_0}^2\Big) +2(M + xm_N)^2\, \nonumber\\
&&\times \mathcal G_1^{\rm iu}\Big(x(x - 1)m_N^2 + (1 - x)M^2 + xm_{qq_0}^2\Big)\bigg\}\,, \rule{1ex}{0ex}
\end{eqnarray}
where $\mathcal G_1^{\rm iu}(\omega)$, $\mathcal G_2^{\rm iu}(\omega)$ are defined in Eqs.\,\eqref{DFG1}, \eqref{DFG2}, respectively, and $e_u = \frac{2}{3}e$.  This expression evaluates to
\begin{align}
  eQ_{p,1} & = D_1 \, e_u \,\mathcal N \nonumber \\
%old  & = 0.0179394\, e_u \,\mathcal N =  0.0119596\,e\,\mathcal N\,.
& = 0.0182622\, e_u \,\mathcal N =  0.0121748\,e\,\mathcal N\,.
  \label{Qp1}
\end{align}

\subsubsection{Diagram 2 -- em}
%%
%%\begin{figure}[h!]
%%  \centering
%%  \vspace{1mm}
%%  \includegraphics[width=0.5\linewidth]{FormFactorFG2.eps}
%%  \caption{Diagram 2: The probe interacts with a quark within the nucleon and the $1^+$ diquark is a bystander.}
%%  \label{DiC2}
%%\end{figure}
The second contribution is almost identical to that depicted in Fig.\,\ref{DiC1}: the only change being that in this instance a $1^+$ diquark is the bystander.  However, owing to isospin symmetry, which we assume herein, and Eq.\,\eqref{aia0}, this term yields
\begin{align}
  eQ_{p,2} & = (2 \,e_d + \,e_u)\,D_{2}^0 \, \mathcal N \nonumber \\
%old  & = (2 \,e_d + \,e_u)\,0.00195023 \, \mathcal N = 0 \,,
%since Eqq1+ did not change, then this should not change.
& = (2 \,e_d + \,e_u)\,0.00195845 \, \mathcal N = 0 \,,
  \label{Qp2}
\end{align}
where $D_{2}^{0}$ is the contribution obtained with a $\{ud\}$-diquark spectator.

\subsubsection{Diagram 3 -- em }
\begin{figure}[t]
  \centering
  \vspace{1mm}
  \includegraphics[width=0.5\linewidth]{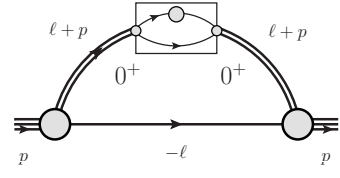}
  \caption{Diagram 3: The probe interacts with the $0^+$ diquark within the proton and the dressed-quark is a bystander.}
  \label{DiC4}
\end{figure}

The third contribution is depicted in Fig.\,\ref{DiC4}, which represents the following expression
{\allowdisplaybreaks
\begin{align}
\label{V4}
  &eQ_{p,3}\,\Lambda^+(p)\gamma_\mu\Lambda^+(p) \nonumber\\
  &= \mathcal N\,\Lambda^+(p)\,\bar{\,\mathcal S\,}\int\frac{d^4\ell}{(2\pi)^4}\,\Delta^{0^+}(\ell + p)\nonumber \\
& \quad \times \mathcal V_{\mu}^0(\ell + p)\Delta^{0^+}(\ell + p)S(-\ell)\,\mathcal S\,\Lambda^+(p)   \\
%  &= -\mathcal N\,\Lambda^+(p)\,s^2\int\frac{d^4\ell}{(2\pi)^4}\frac{1}{(\ell + p)^2 + m_{qq_0}^2}\,\mathcal V_\mu(\ell + p)\,\frac{1}{(\ell + p)^2 + m_{qq_0}^2}\frac{-i\gamma\cdot\ell - M}{\ell^2 + M^2}\,\Lambda^+(p) \nonumber\\
%
  &= -2\,\mathcal N\,\Lambda^+(p)\,s^2\int_0^1 dx\,(1 - x)\int\frac{d^4\ell}{(2\pi)^4}\nonumber \\
&\quad \times \frac{i\gamma\cdot(-\ell + (1 - x)p) - M}{[\ell^2 - x(1 - x)m_N^2 + (1 - x)m_{qq_0}^2 + xM^2]^3}\, \nonumber \\
 & \quad \times \mathcal V_\mu^0(\ell + xp)\,\Lambda^+(p)\,.
\end{align}}

The vertex is given by ($\bar N = 2$)
{\allowdisplaybreaks
\begin{align}
  \mathcal V_\mu^0(P) &= -\bar e_0 \bar N\int\frac{d^4q}{(2\pi)^4}\,\text{tr}\Big\{S(q + P/2)\gamma_\mu S(q + P/2)\nonumber \\
& \quad \times \Gamma_{qq}^{0^+}(P)S(q - P/2)\bar\Gamma_{qq}^{0^+}(-P)\Big\} \\
%
%  &= \bar e_0N_c\int\frac{d^4q}{(2\pi)^4}\,\text{tr}\bigg\{\frac{i\gamma\cdot(q + P/2) - M}{(q + P/2)^2 + M^2}\,\gamma_\mu\,\frac{i\gamma\cdot(q + P/2) - M}{(q + P/2)^2 + M^2}\,\gamma^5\Big(i E_{qq_0} + \frac{1}{M}\,\gamma\cdot P\,F_{qq_0}\Big)\,\frac{i\gamma\cdot(q - P/2) - M}{(q - P/2)^2 + M^2} \nonumber\\
%&\quad\times\gamma^5\Big(i E_{qq_0} - \frac{1}{M}\,\gamma\cdot P\,F_{qq_0}\Big)\bigg\} \nonumber\\
%
  &= 2\bar e_0\bar N\int_0^1dx\,(1 - x)\int\frac{d^4q}{(2\pi)^4}\, \nonumber \\
  & \quad \times \text{tr}\Big\{[i\gamma\cdot(q + xP) - M]\,\gamma_\mu\,[i\gamma\cdot(q + xP) - M]\,\nonumber \\
  & \quad \times \gamma_5\Big(i E_{qq_0} + \frac{1}{M}\,\gamma\cdot P\,F_{qq_0}\Big) \nonumber\\
  &\quad\times[i\gamma\cdot(q + (x-1)P) - M]\,\nonumber \\
 & \quad \times \gamma_5\Big(i E_{qq_0} - \frac{1}{M}\,\gamma\cdot P\,F_{qq_0}\Big)\Big\}\nonumber \\
 & \quad \times \Big(q^2 - x(1 - x)m_{qq_0}^2 + M^2\Big)^{-3}\,,
\end{align}}
\hspace*{-0.5\parindent}where, again, $\bar e_0 = \frac{1}{3}e$; and $P$ is the incoming as well as the outgoing momentum of the $0^+$ diquark, owing to our need to only consider vanishing momentum transfer $Q\to0$, and we choose $P$ to be an on-shell momentum.  Applying the projector in Eq.\,\eqref{eqProjectorCharge} and evaluating the trace, one obtains
\begin{align}
  e Q_{p,3} &= D_3\, \bar e_0 \, \mathcal N \nonumber \\
% old  & = 0.0217886\, \bar e_0\, \mathcal N = 0.00726289\,e\,\mathcal N\,.
& = 0.008733364\, \bar e_0\, \mathcal N = 0.00291112\,e\,\mathcal N\,.
  \label{Qp3}
\end{align}
%0.0217886

\subsubsection{Diagram 4 -- em}
The fourth contribution is almost identical to that depicted in Fig.\,\ref{DiC4}: the only change being that in this instance the $1^+$ diquark is probed, so that one has
{\allowdisplaybreaks
\begin{align}\label{V3}
  &eQ_{p,4}\,\Lambda^+(p)\gamma_\mu\Lambda^+(p) \nonumber\\
  &= \mathcal N\sum_{j\in 0,+}\Lambda^+(p)\,{\mathcal A}_\alpha^j(-p)\int\frac{d^4\ell}{(2\pi)^4}\,\Delta_{\alpha\alpha'}^{1^+}(\ell + p)\mathcal \nonumber \\
  & \quad \times V_{\alpha'\mu\beta'}^j(\ell + p)\Delta_{\beta'\beta}^{1^+}(\ell + p)S(-\ell)\,\mathcal A_\beta^j(p)\,\Lambda^+(p)   \\
%  &= -\mathcal N\sum_{j\in 0,+}\Lambda^+(p)\,\Big(a_1^j\gamma^5\gamma_\alpha + ia_2^j\gamma^5\hat p_\alpha\Big)\int\frac{d^4\ell}{(2\pi)^4}\frac{\mathbb P_{\alpha\alpha'}(\ell + p)}{(\ell + p)^2 + m_{qq_1}^2}\,\mathcal V_{\alpha'\mu\beta'}^j(\ell + p) \nonumber\\
%  &\quad\times\frac{\mathbb P_{\beta'\beta}(\ell + p)}{(\ell + p)^2 + m_{qq_1}^2}\frac{-i\gamma\cdot\ell - M}{\ell^2 + M^2}\Big(a_1^j\gamma_\beta\gamma^5 + ia_2^j\gamma^5\hat p_\beta\Big)\,\Lambda^+(p) \nonumber\\
%
  &= -2\,\mathcal N\sum_{j\in 0,+}\Lambda^+(p)\,\gamma_5\Big(a_1^j\gamma_\alpha + ia_2^j\hat p_\alpha\Big)\int_0^1 dx\,(1 - x) \nonumber \\
  & \quad \times \int\frac{d^4\ell}{(2\pi)^4}\frac{i\gamma\cdot(-\ell + (1 - x)p) - M}{[\ell^2 - x(1 - x)m_N^2 + (1 - x)m_{qq_1}^2 + xM^2]^3} \nonumber\\
  &\quad\times\mathbb P_{\alpha\alpha'}(\ell + xp)\,\mathcal V_{\alpha'\mu\beta'}^j(\ell + xp)\,\mathbb P_{\beta'\beta}(\ell + xp)\nonumber \\
  & \quad \times \Big(a_1^j\gamma_\beta + ia_2^j\hat p_\beta\Big)\gamma_5\,\Lambda^+(p)\,.
\end{align}}

The vertex is ($\bar N = 2$)
{\allowdisplaybreaks
\begin{align}
\mathcal V_{\alpha\mu\beta}^j&(P)  = -\bar e_j \bar N \int\frac{d^4q}{(2\pi)^4}\,\text{tr}\Big\{S(q + P/2)\gamma_\mu S(q + P/2) \nonumber \\
 & \quad \times \Gamma_{qq\beta }^{1^+}(P)S(q - P/2)\bar\Gamma_{qq\alpha }^{1^+}(-P)\Big\} \\
%  &\hspace{-5mm}= \bar e_jN_c\int\frac{d^4q}{(2\pi)^4}\,\text{tr}\bigg\{\frac{i\gamma\cdot(q + P/2) - M}{(q + P/2)^2 + M^2}\,\gamma_\mu\,\frac{i\gamma\cdot(q + P/2) - M}{(q + P/2)^2 + M^2}\,iE_{qq_1}\gamma_\beta^T(P)\,\frac{i\gamma\cdot(q - P/2) - M}{(q - P/2)^2 + M^2}\,iE_{qq_1}\gamma_\alpha^T(P)\bigg\} \nonumber\\
  &= -2\bar e_j \bar N E_{qq_1}^2\int_0^1dx\,(1 - x)\int\frac{d^4q}{(2\pi)^4}\,\nonumber \\
  & \quad \times \text{tr}\{[i\gamma\cdot(q + xP) - M]\,\gamma_\mu\,[i\gamma\cdot(q + xP) - M]\,\nonumber \\
  & \quad \times \gamma_\beta^T(P)\,[i\gamma\cdot(q + (x-1)P) - M]\,\gamma_\alpha^T(P)\} \nonumber \\
 &\quad \times [q^2 - x(1 - x)m_{qq_1}^2 + M^2]^{-3}\,,
\end{align}}
\hspace*{-0.5\parindent}where, as noted above, $\bar e_0 = \frac{1}{3}e$ and $\bar e_+ = \frac{4}{3}e$, and $P$ is the incoming as well as outgoing momentum of the $1^+$ diquark.  Applying the projector in Eq.\,\eqref{eqProjectorCharge} and evaluating the trace, one obtains
\begin{align}
%  \rule{-1.9ex}{0ex}
  e&Q_{p,4} = (2 \, \bar e_{+} + \bar e_0 ) \, D_4^0 \, \mathcal N \nonumber \\
%old  &= (2 \, \bar e_{+} + \bar e_0 )\,0.0009038\, \mathcal N = 0.002711\,e\,\mathcal N\,,
  &= (2 \, \bar e_{+} + \bar e_0 )\,0.00090133\, \mathcal N = 0.002704\,e\,\mathcal N\,,
  \label{Qp4}
\end{align}
where $D_4^0$ is the contribution from the $\{ud\}$-diquark.
% 0.00272848

\subsubsection{Diagram 5 -- em}
\begin{figure}[t]
  \centering
  \vspace{1mm}
  \includegraphics[width=0.5\linewidth]{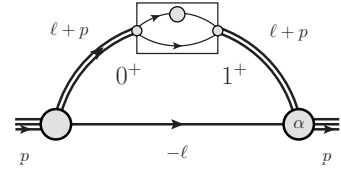}
  \caption{Diagram 5: The probe is absorbed by a $0^+$-diquark, which is thereby transformed into a $1^+$ diquark.}
  \label{DiC6}
\end{figure}

This contribution is depicted in Fig.\,\ref{DiC6}.  In this case
\begin{align}\label{V5}
  Q_{p,5}e\,\Lambda^+(p)\gamma_\mu\Lambda^+(p) = 0\,,
\end{align}
because the vertex vanishes at zero momentum transfer; i.e.,
\begin{align}
  \mathcal V_{\mu\alpha} = 0\,.
\end{align}
Consequently
\begin{equation}
Q_{p,5} = 0\,.
\end{equation}

\subsubsection{Diagram 6 -- em}
This is the conjugate contribution to that depicted in Fig.\,\ref{DiC6}; namely, a $1^+$ diquark absorbs the probe and is thereby transformed into a $0^+$ diquark.  In a symmetry preserving treatment of any reasonable interaction, this contribution is identical to that produced by Diagram~5.

\subsubsection{Current Conservation}
\label{sscc}
If a truly Poincar\'e invariant regularisation is employed, then one has Ward identities relating the charges in Eqs.\,\eqref{Qp1}, \eqref{Qp4} and \eqref{Qp2}, \eqref{Qp3}
\begin{equation}
D_1=D_3\,, \; D_2^0=D_4^0\,,
\end{equation}
which ensure: simple additivity of the quark and diquark electric charges, and thereby guarantee a unit-charge isospin=$(+1/2)$ baryon through a single rescaling factor \cite{Oettel:1999gc}; and a neutral isospin=$(-1/2)$ baryon without fine tuning.  Owing to the cutoffs we have introduced, however, these identities are violated: Eq.\,\eqref{Qp1} cf.\ \eqref{Qp3}, Eq.\,\eqref{Qp2} cf.\ \eqref{Qp4}.  Following Ref.\,\cite{Wilson:2011aa}, we ameliorate this flaw by enforcing the Ward identities:
\begin{subequations}
\label{Dvalues}
\begin{align}
% old D_{1,3} &\to D_{\overline{13}} = (D_1+D_3)/2 = 0.019864 \,,\; \\
D_{1,3} &\to D_{\overline{13}} = (D_1+D_3)/2 = 0.01350 \,,\; \\
% old D_{2,4} &\to D_{\overline{24}} = (D_2+D_4)/2 = 0.001427\,.
D_{2,4} &\to D_{\overline{24}} = 3(D_2^0+D_4^0)/2 = 0.00429\,.
\end{align}
\end{subequations}
This corresponds to introducing a rescaling factor for each of the diagrams involved: $D_i\to \kappa_i D_i$, $\kappa_{1,3} = D_{\overline{13}}/D_{1,3}$, $\kappa_{2,4} = D_{\overline{24}}/D_{2,4}$.  Diagrams 5 and 6 are unaffected because they are equal and do not contribute to a baryon's charge.

\subsubsection{Canonical Normalisation}
The results computed from all diagrams considered in connection with the proton's charge are collected in Table~\ref{VT}.  As noted above, the canonical normalisation is fixed by requiring
\begin{equation}
Q_p = \sum_{i=1}^6\, Q_{p,i} = 1\,,
\end{equation}
from which it follows that
\begin{equation}
\label{normalisation}
% old \mathcal N = \frac{1}{0.0241} = 41.42\,.
\mathcal N = \frac{1}{0.01777} = 56.27\,.
\end{equation}

\begin{table}[t!]
\begin{center}
\begin{tabular}{|c|c|c|}
\hline
 & $Q_{p,i}/\mathcal N$ &  $Q^\kappa_{p,i}/\mathcal N$ \\
\hline\hline
Diagram 1 & $0.01217$ & 0.0090  \\
\hline
Diagram 2 & $0$ & 0  \\
\hline
Diagram 3 & $0.00291$ & $0.00450$\\
\hline
Diagram 4 & $0.00270$ & $0.00426$ \\
\hline
Diagram 5 & $0$ &   \\
\hline
Diagram 6 & $0$ &   \\
\hline\hline
Sum & $0.0178$ & $0.0178$ \\
\hline
\end{tabular}
\end{center}
\caption{\label{VT}
Column 1: Summary of the results computed from all diagrams considered in connection with the proton's charge.  Column 2: Results scaled as described in Sec.\,\ref{sscc}.}
\end{table}

\subsection{Scalar Current}
\label{AppScalarCurrent}
When computing the scalar charge of any hadron, one must employ the dressed-quark-scalar vertex.  That vertex, too, is obtained by solving an inhomogeneous Bethe-Salpeter equation: in this case, the unrenormalised form is determined by the inhomogeneous term ${\mathbb 1}$.  The complete solution for the contact-interaction's scalar vertex in RL truncation can be found in Refs.\,\cite{Chen:2012txa}, and at $Q^2=0$ this yields:
%% 1:37415
\begin{equation}
 \mathcal V^q_{\mathbb 1} = \frac{1}{\displaystyle 1 + \frac{4\alpha_\text{IR}}{3\pi m_G^2}\Big(2\,\mathcal C_1^\text{iu}(M^2) - \mathcal C^\text{iu}(M^2)\Big)}\,{\mathbb 1} = 1.37\,{\mathbb 1}\,,
 \label{ScalarVertexV}
\end{equation}
where $M$ is the dressed-quark mass in Eq.\,\eqref{dressedMvalue}.

As a check on this result, we note again that since the vertex is only required at $Q^2=0$, one can appeal to a Ward identity \cite{Chang:2008ec}, which takes the form
\begin{equation}
{\mathcal V}_{\mathbb 1}(Q) \stackrel{Q^2=0}{=} {\mathbb 1} \, \frac{\partial M}{\partial m}\,
\end{equation}
when the contact interaction is used.
%%%1.37203
Employing the results from which Ref.\,\cite{Roberts:2011cf} was prepared, this expression, too, yields the numerical value in Eq.\,\eqref{ScalarVertexV}.

The nucleon's scalar charge is also known as the nucleon $\sigma$-term; and using our implementation of the contact interaction, one need consider only relevant analogues of the six diagrams described explicitly in App.\,\ref{emApp}.  In this case, Diagrams~1--4 provide a nonzero contribution and the complete result is obtained from the sum.

\subsubsection{Diagram 1 -- scalar}
This is the contribution produced by the scalar probe interacting with a the dressed-quark whilst the $0^+$ $[ud]$-diquark is a spectator:
\begin{align}
  \hat\sigma_{q,1}&\,\Lambda^+(p) {\mathbb 1}\Lambda^+(p) =
  {\mathcal N}_1^\kappa\,\Lambda^+(p)\,\bar{\,\mathcal S\,}\int\frac{d^4\ell}{(2\pi)^4}\,S(\ell + p) \nonumber\\
 & \quad \times  \mathcal V^q_{\mathbb 1} S(\ell + p)\Delta^{0^+}(-\ell)\,\mathcal S\,\Lambda^+(p) \\
%  &= \mathcal N\,\Lambda^+(p)\,s^2\int\frac{d^4\ell}{(2\pi)^4}\frac{i\gamma\cdot(\ell + p) - M}{(\ell + p)^2 + M^2}\,\mathcal V^q\,\frac{i\gamma\cdot(\ell + p) - M}{(\ell + p)^2 + M^2}\frac{1}{\ell^2 + m_{qq_0}^2}\,\Lambda^+(p) \nonumber\\
%
  &= 2\,{\mathcal N}_1^\kappa\,\Lambda^+(p)\,s^2\int_0^1 dx\,(1 - x)\int\frac{d^4\ell}{(2\pi)^4} \nonumber \\
& \; \times \frac{\{i\gamma\cdot(\ell + x p) - M\}\,\mathcal V_{\mathbb 1}^q\,\{i\gamma\cdot(\ell + x p) - M\}}{[\ell^2 - x(1 - x)m_N^2 + (1 - x)M^2 + xm_{qq_0}^2]^3}\,\Lambda^+(p)\,,
\end{align}
where ${\mathcal N}_1^\kappa = \kappa_1 {\mathcal N}$, with $\kappa_1$ defined in connection with Eqs.\,\eqref{Dvalues}, ${\mathcal N}$ given in Eq.\,\eqref{normalisation}.  Applying the projector
\begin{equation}
\label{projectorS}
  \mathcal P = \frac{1}{2}\,{\mathbb 1}\,,
\end{equation}
and evaluating the trace, one obtains
\begin{equation}
%  \hat\sigma_{u,1} = \sigma_{q,1} = 0.330\,, \; ... unscaled
  \hat\sigma_{u,1} = \hat\sigma_{q,1} = 0.309\,, \; %scaled
  \hat\sigma_{d,1} = 0\,.
\end{equation}
% (NF/NFO) 0.330*0.019864/0.0179394
It was plain from the outset that this diagram would only produce a contribution to $\hat\sigma_{u,1}$ because the $d$-quark is sequestered inside the scalar diquark.

\subsubsection{Diagram 2 -- scalar}
In this case we have the scalar probe interacting with the dressed-quark and the $1^+$ diquarks being spectators:
\begin{align}
  &\hat\sigma_{q_{j,2}}\,\Lambda^+(p){\mathbb 1}\Lambda^+(p) \nonumber\\
  &={\mathcal N}_2^\kappa\,\Lambda^+(p)\,{\mathcal A}_\alpha^j(-p)\int\frac{d^4\ell}{(2\pi)^4}\,S(\ell + p)\mathcal V^q_{\mathbb 1}\nonumber\\
& \quad \times S(\ell + p)\Delta_{\alpha\beta}^{1^+}(-\ell)\,\mathcal A_\beta^j(p)\,\Lambda^+(p)  \\
%
%&= \mathcal N\,\Lambda^+(p)\,\Big(a_1^j\gamma^5\gamma_\alpha + ia_2^j\gamma^5\hat p_\alpha\Big)\int\frac{d^4\ell}{(2\pi)^4} \! \frac{i\gamma\cdot(\ell + p) - M}{(\ell + p)^2 + M^2}\,\mathcal V^q\,\frac{i\gamma\cdot(\ell + p) - M}{(\ell + p)^2 + M^2}\frac{\mathbb P_{\alpha\beta}(\ell)}{\ell^2 + m_{qq_1}^2} \nonumber\\
%&\quad\times\Big(a_1^j\gamma_\beta\gamma^5 + ia_2^j\gamma^5\hat p_\beta\Big)\,\Lambda^+(p) \nonumber\\
%
  &= 2\,{\mathcal N}_2^\kappa\,\Lambda^+(p)\,\gamma_5\Big(a_1^j\gamma_\alpha + ia_2^j\hat p_\alpha\Big)\!\!\int_0^1 dx\,(1 - x)\!\!\int\frac{d^4\ell}{(2\pi)^4}\nonumber\\
&\quad \times \frac{\{i\gamma\cdot(\ell + x p) - M\}\,\mathcal V^q_{\mathbb 1}\,\{i\gamma\cdot(\ell + x p) - M\}}{[\ell^2 - x(1 - x)m_N^2 + (1 - x)M^2 + xm_{qq^1}^2]^3} \nonumber\\
  &\quad\times\mathbb P_{\alpha\beta}(\ell - (1 - x)p)\!\Big(a_1^j\gamma_\beta + ia_2^j\hat p_\beta\Big)\gamma_5\,\Lambda^+(p). \hspace*{-1ex}
\end{align}
Applying the projector in Eq.\,\eqref{projectorS} and evaluating the trace, one finds, owing to Eq.\,\eqref{aia0},
\begin{equation}
  \hat\sigma_{u,2} = \hat\sigma_{q_{0,2}} = 0.0318\,, \;
  \hat\sigma_{d,2} = \hat\sigma_{q_{+,2}} = 0.0636 = 2 \hat\sigma_{u,2}\,.
\end{equation}
% (NF/NFO) 0.035*0.001427/0.00195023 = 0.0232718

\subsubsection{Diagram 3 -- scalar}
The third diagram describes the scalar probe interacting with the $0^+$ $[ud]$-diquark and the dressed-quark acting merely as an onlooker:
{\allowdisplaybreaks
\begin{align}\label{S4}
  &\hat\sigma_{q,3}\,\Lambda^+(p){\mathbb 1}\Lambda^+(p) =
  {\mathcal N}_3^\kappa\,\Lambda^+(p)\,\bar{\,\mathcal S\,}\int\frac{d^4\ell}{(2\pi)^4}\,\Delta^{0^+}(\ell + p)\nonumber\\
  &\quad \times \mathcal V_{\mathbb 1}^0(\ell+p)\Delta^{0^+}(\ell + p)S(-\ell)\,\mathcal S\,\Lambda^+(p)   \\
%%
%  &= -\mathcal N\,\Lambda^+(p)\,s^2\int\frac{d^4\ell}{(2\pi)^4}\frac{1}{(\ell + p)^2 + m_{qq_0}^2}\,\mathcal V\,\frac{1}{(\ell + p)^2 + m_{qq_0}^2}\frac{-i\gamma\cdot\ell - M}{\ell^2 + M^2}\,\Lambda^+(p) \nonumber\\
%%
  &= -2\,{\mathcal N}_3^\kappa\,s^2\int_0^1 dx\,(1 - x)\int\frac{d^4\ell}{(2\pi)^4}\Lambda^+(p)\nonumber\\
    &  \times \frac{[i\gamma\cdot(-\ell + (1 - x)p) - M]\mathcal V^0_{\mathbb 1}(\ell+xp)\,\Lambda^+(p)}{[\ell^2 - x(1 - x)m_N^2 + (1 - x)m_{qq^0}^2 + xM^2]^3}\,.
\end{align}}
\hspace*{-0.5\parindent}The vertex is given by ($\bar N = 2$)
{\allowdisplaybreaks
\begin{align}
  &\mathcal V^0_{\mathbb 1}(P)= -2\bar N\int\frac{d^4q}{(2\pi)^4}\,\text{tr}\Big\{S(q + P/2)\mathcal V^q_{\mathbb 1} S(q + P/2)\nonumber\\
    & \quad \times \Gamma_{qq}^{0^+}(P)S(q - P/2)\bar\Gamma_{qq}^{0^+}(-P)\Big\} \\
%%
%&= 2N_c\int\frac{d^4q}{(2\pi)^4}\,\text{tr}\bigg\{\frac{i\gamma\cdot(q + P/2) - M}{(q + P/2)^2 + M^2}\,\mathcal V^q\,\frac{i\gamma\cdot(q + P/2) - M}{(q + P/2)^2 + M^2}\,\gamma_5\Big(i E_{qq_0} + \frac{1}{M}\,\gamma\cdot P\,F_{qq_0}\Big)\,\frac{i\gamma\cdot(q - P/2) - M}{(q - P/2)^2 + M^2} \nonumber\\
%  &\quad\times\gamma^5\Big(i E_{qq_0} - \frac{1}{M}\,\gamma\cdot P\,F_{qq_0}\Big)\bigg\} \nonumber\\
%%
  &= 4\bar N\int_0^1dx\,(1 - x)\int\frac{d^4q}{(2\pi)^4}\,\text{tr}\Big\{[i\gamma\cdot(q + xP) - M]\,\nonumber\\
&\quad \times \mathcal V^q_{\mathbb 1}\,[i\gamma\cdot(q + xP) - M]\,\gamma_5\Big(i E_{qq_0} + \frac{1}{M}\,\gamma\cdot P\,F_{qq_0}\Big) \nonumber\\
  &\quad\times[i\gamma\cdot(q + (x-1)P) - M]\,\gamma_5\Big(i E_{qq_0} \nonumber\\
&\quad - \frac{1}{M}\,\gamma\cdot P\,F_{qq_0}\Big)\Big\}\Big(q^2 - x(1 - x)m_{qq_0}^2 + M^2\Big)^{-3}\,.
\end{align}}

Applying the projector in Eq.\,\eqref{projectorS} and evaluating the trace, one obtains
\begin{equation}
  \hat\sigma_{u,3} = \frac{\hat\sigma_{q,3}}{2} = 1.0419 = \hat\sigma_{d,3} \,.
\end{equation}
% (NF/NFO) 1.364*0.019864/0.0217886

\subsubsection{Diagram 4 -- scalar}
The fourth diagram describes the scalar probe interacting with a $1^+$ $\{uu\}$- or $\{ud\}$-diquark where the dressed-quark acts merely as an onlooker:
\begin{align}\label{S3}
  &\hat\sigma_{q_{j,4}}\,\Lambda^+(p){\mathbb 1}\Lambda^+(p) \nonumber\\
  &= {\mathcal N}_4^\kappa\,\Lambda^+(p)\,{\mathcal A}_\alpha^j(-p)\int\frac{d^4\ell}{(2\pi)^4}\,\Delta_{\alpha\alpha'}^{1^+}(\ell + p)\mathcal V_{\alpha'\beta'}^{\mathbb 1}(\ell + p)\nonumber\\
  & \quad \times \Delta_{\beta'\beta}^{1^+}(\ell + p)S(-\ell)\,\mathcal A_\beta^j(p)\,\Lambda^+(p)   \\
%%
%  &= -\mathcal N\,\Lambda^+(p)\,\Big(a_1^j\gamma^5\gamma_\alpha + ia_2^j\gamma^5\hat p_\alpha\Big)\int\frac{d^4\ell}{(2\pi)^4}\frac{\mathbb P_{\alpha\alpha'}(\ell + p)}{(\ell + p)^2 + m_{qq_1}^2}\,\mathcal V_{\alpha'\beta'}(\ell + p)\,\frac{\mathbb P_{\beta'\beta}(\ell + p)}{(\ell + p)^2 + m_{qq_1}^2}\frac{-i\gamma\cdot\ell - M}{\ell^2 + M^2} \nonumber\\
%  &\quad\times\Big(a_1^j\gamma_\beta\gamma^5 + ia_2^j\gamma^5\hat p_\beta\Big)\,\Lambda^+(p) \nonumber\\
%%
  &= -2\,{\mathcal N}_4^\kappa\,\Lambda^+(p)\,\gamma_5\Big(a_1^j\gamma_\alpha + ia_2^j\hat p_\alpha\Big)\int_0^1 dx\,(1 - x)\nonumber\\
  & \quad \times \int\frac{d^4\ell}{(2\pi)^4}\frac{i\gamma\cdot(-\ell + (1 - x)p) - M}{[\ell^2 - x(1 - x)m_N^2 + (1 - x)m_{qq_1}^2 + xM^2]^3} \nonumber\\
  & \quad\times\mathbb P_{\alpha\alpha'}(\ell + xp)\,\mathcal V^{\mathbb 1}_{\alpha'\beta'}(\ell + xp)\,\mathbb P_{\beta'\beta}(\ell + xp)\nonumber\\
 & \quad \times \Big(a_1^j\gamma_\beta + ia_2^j\hat p_\beta\Big)\gamma_5\,\Lambda^+(p)\,.
\end{align}
The vertex is given by ($\bar N = 2$)
{\allowdisplaybreaks
\begin{align}
  &\mathcal V^{\mathbb 1}_{\alpha\beta}(P) = -2\bar N\int\frac{d^4q}{(2\pi)^4}\,\text{tr}\Big\{S(q + P/2)\mathcal V_{\mathbb 1}^q S(q + P/2)\nonumber\\
  & \quad\times \Gamma_{qq \beta}^{1^+}(P)S(q - P/2) \bar\Gamma^{1^+}_{qq \alpha}(-P)\Big\} \\
%  &\hspace{-5mm}= 2N_c\int\frac{d^4q}{(2\pi)^4}\,\text{tr}\bigg\{\frac{i\gamma\cdot(q + P/2) - M}{(q + P/2)^2 + M^2}\,\mathcal V^q\,\frac{i\gamma\cdot(q + P/2) - M}{(q + P/2)^2 + M^2}\,iE_{qq_1}\gamma_\beta^T(P)\,\frac{i\gamma\cdot(q - P/2) - M}{(q - P/2)^2 + M^2}\,iE_{qq_1}\gamma_\alpha^T(P)\bigg\} \nonumber\\
  &= -4 \bar N E_{qq_1}^2\int_0^1dx\,(1 - x)\int\frac{d^4q}{(2\pi)^4}\,\text{tr}\{[i\gamma\cdot(q + xP) \nonumber \\
  & \quad - M]\,\mathcal V^q_{\mathbb 1}\,[i\gamma\cdot(q + xP) - M]\,\gamma_\beta^T\,[i\gamma\cdot(q + (x-1)P) \nonumber \\
  & \quad - M]\,\gamma_\alpha^T\}[q^2 - x(1 - x)m_{qq_1}^2 + M^2]^{-3} \\
  &\to 16 M \bar N E_{qq_1}^2\mathcal V^q_{\mathbb 1}
  \mathbb P_{\alpha\beta}(P)\int_0^1dx\,(1 - x)\nonumber\\
  & \quad \times \Big(M^2 - x(x - 2)m_{qq_1}^2\Big)\,
%  \mathcal G_1^{\rm iu}(\omega)\Big|_{\omega = x(x - 1)m_{qq_1}^2 + M^2}\,,
  \mathcal G_1^{\rm iu}\Big(x(x - 1)m_{qq_1}^2 + M^2\Big)\,,
\end{align}}
\hspace*{-0.5\parindent}where $P$ is again both the incoming and outgoing momentum of the $1^+$ diquark.

Applying the projector in Eq.\,\eqref{projectorS} and evaluating the trace, one finds \begin{equation}
  \hat\sigma_{u,4} = \frac{\hat\sigma_{q_{0,4}}}{2} + \hat\sigma_{q_{+,4}} = 0.465\,, \;  \hat\sigma_{d,4} = \frac{\hat\sigma_{q_{0,4}}}{2} = 0.0938\,.
\end{equation}
%(NF/NFO) 0.0482 k24/0.00090376=0.0691575
% (NF/NFO) (0.239 - 0.0482) k24/0.00090376 + (NF/NFO) 0.0482 k24/ 0.00090376=0.342918

\subsubsection{Proton $\mathbf \sigma$-term}
The results obtained from all diagrams considered in connection with the proton's scalar charge are collected in Table~\ref{Nsigma}.  The proton $\sigma$-term is
\begin{equation}
\label{eqNsigma}
\sigma_N = m \sum_{i=1}^6 [ \hat\sigma_{u,i} + \hat\sigma_{d,i} ] =
%%...unscaled 24.2\,{\rm MeV}.
21.33\,{\rm MeV}.
\end{equation}
In the isospin symmetric limit, the neutron $\sigma$-term is identical.

\begin{table}[t!]
\begin{center}
\begin{tabular}{|c|c|c|l|}
\hline
 & $\hat\sigma_u$ & $\hat\sigma_d$ &  $\sigma$ [MeV]  \\
\hline\hline
Diagram 1 & $0.309$ & $0$ & $ \phantom{1}2.163$  \\
\hline
Diagram 2 & $0.032$ & $0.063$ & $\phantom{1}0.666$  \\
\hline
Diagram 3 & $1.042$ & $1.042$ & $14.587$  \\
\hline
Diagram 4 & $0.465$ & $0.094$ & $\phantom{1}3.914$  \\
\hline
Diagram 5 & $0$ & $0$ & $\phantom{1}0$  \\
\hline
Diagram 6 & $0$ & $0$ & $\phantom{1}0$   \\
\hline\hline
Total Result & $1.85$ & $1.20$ & $21.33$  \\
\hline
%FHT & & & $26.8$  \\
%\hline
\end{tabular}
\caption{\label{Nsigma}
Summary of the results computed from all diagrams considered in connection with the proton's scalar charge.}
\end{center}
\end{table}

\subsection{Tensor Current}
\label{AppTensor}
When computing the tensor charge of any hadron, one must employ the dressed-quark-tensor vertex.  However, as explained elsewhere \cite{Roberts:2011cf}, any dressing of the tensor vertex must depend linearly on the relative momentum \cite{LlewellynSmith:1969az} and such dependence is impossible using a symmetry-preserving regularisation of a vector$\,\otimes\,$vector contact interaction. Hence, in our case, the quark-tensor vertex is unmodified from its bare form; viz.,
\begin{equation}
\label{DressedQT}
\mathcal V_{\mu\nu}^q = \sigma_{\mu\nu}\,.
\end{equation}

Naturally, when computing the proton's tensor charge using our implementation of the contact interaction, one need only consider relevant analogues of the six diagrams described explicitly in App.\,\ref{emApp}.  In this case, Diagrams~1,2,4,5,6 provide nonzero contributions.  Diagram~3 yields zero because Poincar\'e invariance entails that a scalar diquark cannot possess a tensor charge.

\subsubsection{Diagram 1 -- tensor}
As usual, we first consider the case of the tensor probe interacting with the dressed-quark and the $0^+$ $[ud]$-diquark being a spectator:
{\allowdisplaybreaks
\begin{align}
  \delta_1 q &\,\Lambda^+(p)\sigma_{\mu\nu}\Lambda^+(p) =
  {\mathcal N}_1^\kappa\,\Lambda^+(p)\,\bar{\,\mathcal S\,}\int\frac{d^4\ell}{(2\pi)^4}\,S(\ell + p)\sigma_{\mu\nu}\nonumber \\
  &\quad \times S(\ell + p)\Delta^{0^+}(-\ell)\,\mathcal S\,\Lambda^+(p) \\
%%
%  &= \mathcal N\,\Lambda^+(p)\,s^2\int\frac{d^4\ell}{(2\pi)^4}\frac{i\gamma\cdot(\ell + p) - M}{(\ell + p)^2 + M^2}\,\sigma_{\mu\nu}\,\frac{i\gamma\cdot(\ell + p) - M}{(\ell + p)^2 + M^2}\frac{1}{\ell^2 + m_{qq_0}^2}\,\Lambda^+(p) \nonumber\\
  &= 2\,\mathcal N\,s^2\int_0^1 dx\,(1 - x)\int\frac{d^4\ell}{(2\pi)^4}
  \Lambda^+(p)\,\{i\gamma\cdot(\ell + x p)\nonumber \\
  & \quad - M\}\,\sigma_{\mu\nu}\,\{i\gamma\cdot(\ell + x p) - M\}\Lambda^+(p)\,\nonumber\\
  &\quad \times [\ell^2 - x(1 - x)m_N^2 + (1 - x)M^2 + xm_{qq_0}^2]^{-3}\,.
\end{align}}
\hspace*{-0.5\parindent}Applying the projector
\begin{align}
\label{ProjectorTensor}
  \mathcal P_{\mu\nu} = \frac{1}{12}\,\sigma_{\mu\nu}\,,
\end{align}
and evaluating the trace, one obtains
\begin{align}
  \delta_1 q &= 2\,{\mathcal N}_1\,s^2\int_0^1 dx\,(1 - x)\int\frac{d^4\ell}{(2\pi)^4}\nonumber\\
  & \quad \times \frac{(M + xm_N)^2}{[\ell^2 - x(1 - x)m_N^2 + (1 - x)M^2 + xm_{qq_0}^2]^3} \\
  &\to 2\,\mathcal N\,s^2\int_0^1dx\,(1 - x)(M + xm_N)^2\,\nonumber\\
  & \quad \times\mathcal G_1^{\rm iu}\Big(x(x - 1)m_N^2 + (1 - x)M^2 + xm_{qq_0}^2\Big) \,,
  \label{D1Tensor}
\end{align}
where $\mathcal G_1^{\rm iu}(\omega)$ is defined in Eq.~(\ref{DFG1}). As a result we find
\begin{equation}
  \delta_{T1} u = \delta_1 q = 0.581\,, \quad  % 0.628871 %0.581316
  \delta_{T1} d = 0\,.
\end{equation}

\subsubsection{Diagram 2 -- tensor}
When the tensor probe interacts with the dressed-quark and the $1^+$ diquarks are spectators, one has
{\allowdisplaybreaks
\begin{align}
  &\delta_2 q_j\,\Lambda^+(p)\sigma_{\mu\nu}\Lambda^+(p) \nonumber\\
  &={\mathcal N}_2^\kappa\,\Lambda^+(p)\,{\mathcal A}_\alpha^j(-p)\int\frac{d^4\ell}{(2\pi)^4}\,S(\ell + p)\sigma_{\mu\nu}\nonumber\\
  &\quad \times S(\ell + p)\,\Delta_{\alpha\beta}^{1^+}(-\ell)\,\mathcal A_\beta^j(p)\,\Lambda^+(p)  \\
%  &= \mathcal N\,\Lambda^+(p)\,\Big(a_1^j\gamma^5\gamma_\alpha + ia_2^j\gamma^5\hat p_\alpha\Big)\int\frac{d^4\ell}{(2\pi)^4}\frac{i\gamma\cdot(\ell + p) - M}{(\ell + p)^2 + M^2}\,\sigma_{\mu\nu}\,\frac{i\gamma\cdot(\ell + p) - M}{(\ell + p)^2 + M^2}\frac{\mathbb P_{\alpha\beta}(\ell)}{\ell^2 + m_{qq_1}^2} \nonumber\\
%  &\quad\times\Big(a_1^j\gamma_\beta\gamma^5 + ia_2^j\gamma^5\hat p_\beta\Big)\,\Lambda^+(p) \nonumber\\
%%
  &= 2\,{\mathcal N}_2^\kappa\,\Lambda^+(p)\,\gamma_5\Big(a_1^j\gamma_\alpha + ia_2^j\hat p_\alpha\Big)\int_0^1 dx\,(1 - x)\int\frac{d^4\ell}{(2\pi)^4}\nonumber\\
  &\quad\times \frac{\{i\gamma\cdot(\ell + x p) - M\}\,\sigma_{\mu\nu}\,\{i\gamma\cdot(\ell + x p) - M\}}{[\ell^2 - x(1 - x)m_N^2 + (1 - x)M^2 + xm_{qq_1}^2]^3} \nonumber\\
  &\quad\times\mathbb P_{\alpha\beta}(\ell - (1 - x)p)\Big(a_1^j\gamma_\beta + ia_2^j\hat p_\beta\Big)\gamma_5\Lambda^+(p).
\end{align}}
Applying the projector in Eq.\,\eqref{ProjectorTensor} and evaluating the resulting trace, one finds, owing to Eq.\,\eqref{aia0}:
%\begin{equation}
%  \delta_{T2} u = \delta_2 q_0 = -0.0200217\,, \nonumber\\
%  \delta_{T2} d = \delta_2 q_+ = -0.0396537\,.
%\end{equation}
%0.0397836
%0.0198918
\begin{equation}
\delta_{T2} d = \delta_2 q_+ = 2 \delta_2 q_0 = -0.0359 = 2 \delta_{T2} u \,.
\end{equation}
%% 0.0398 -> 0.0264632
%% -0.0179292*2=-0.0358583

\subsubsection{Diagram 4 -- tensor}
The next nonzero contribution arises from the tensor probe interacting with a $1^+$ $\{uu\}$- or $\{ud\}$-diquark where the dressed-quark acts merely as an onlooker:
{\allowdisplaybreaks
\begin{align}\label{T3}
  &\delta_4 q_j\,\Lambda^+(p)\sigma_{\mu\nu}\Lambda^+(p) \nonumber\\
  &= {\mathcal N}_4^\kappa\,\Lambda^+(p)\,{\mathcal A}_\alpha^j(-p)\int\frac{d^4\ell}{(2\pi)^4}\,\Delta_{\alpha\alpha'}^{1^+}(\ell + p)\mathcal V^2_{\alpha'\mu\nu\beta'}(\ell + p)\nonumber\\
  & \quad\times \Delta_{\beta'\beta}^{1^+}(\ell + p)S(-\ell)\,\mathcal A_\beta^j(p)\,\Lambda^+(p)   \\
%%
%  &= -\mathcal N\,\Lambda^+(p)\,\Big(a_1^j\gamma^5\gamma_\alpha + ia_2^j\gamma^5\hat p_\alpha\Big)\int\frac{d^4\ell}{(2\pi)^4}\frac{\mathbb P_{\alpha\alpha'}(\ell + p)}{(\ell + p)^2 + m_{qq_1}^2}\,\mathcal V_{\alpha'\mu\nu\beta'}(\ell + p)\,\frac{\mathbb P_{\beta'\beta}(\ell + p)}{(\ell + p)^2 + m_{qq_1}^2}\frac{-i\gamma\cdot\ell - M}{\ell^2 + M^2} \nonumber\\
%  &\quad\times\Big(a_1^j\gamma_\beta\gamma^5 + ia_2^j\gamma^5\hat p_\beta\Big)\,\Lambda^+(p) \nonumber\\
%%
  &= -2\,{\mathcal N}_4^\kappa\,\Lambda^+(p)\,\gamma_5\Big(a_1^j\gamma_\alpha + ia_2^j\hat p_\alpha\Big)\int_0^1 dx\,(1 - x)\int\frac{d^4\ell}{(2\pi)^4}\nonumber\\
  & \quad\times \frac{i\gamma\cdot(-\ell + (1 - x)p) - M}{[\ell^2 - x(1 - x)m_N^2 + (1 - x)m_{qq_1}^2 + xM^2]^3} \nonumber\\
  &\quad\times\mathbb P_{\alpha\alpha'}(\ell + xp)\,\mathcal V^2_{\alpha'\mu\nu\beta'}(\ell + xp)\,\nonumber\\
  &\quad\times \mathbb P_{\beta'\beta}(\ell + xp)\Big(a_1^j\gamma_\beta + ia_2^j\hat p_\beta\Big)\gamma_5\,\Lambda^+(p)\,.
\end{align}}

The vertex is ($\bar N = 2$)
{\allowdisplaybreaks
\begin{align}
  &\mathcal V^2_{\alpha\mu\nu\beta}(P) = -2 \bar N \int\frac{d^4q}{(2\pi)^4}\,\text{tr}\Big\{S(q + P/2)\sigma_{\mu\nu} S(q + P/2)\nonumber\\
  &\quad\times \Gamma_{qq\beta}^{1^+}(P) S(q - P/2) \bar\Gamma_{qq\alpha}^{1^+}(-P)\Big\} \\
%  &\hspace{-5mm}= 2N_c\int\frac{d^4q}{(2\pi)^4}\,\text{tr}\bigg\{\frac{i\gamma\cdot(q + P/2) - M}{(q + P/2)^2 + M^2}\,\sigma_{\mu\nu}\,\frac{i\gamma\cdot(q + P/2) - M}{(q + P/2)^2 + M^2}\,iE_{qq_1}\gamma_\beta^T(P)\,\frac{i\gamma\cdot(q - P/2) - M}{(q - P/2)^2 + M^2}\,iE_{qq_1}\gamma_\alpha^T(P)\bigg\} \nonumber\\
  &= -4 \bar N E_{qq_1}^2\int_0^1dx\,(1 - x)\int\frac{d^4q}{(2\pi)^4}\,
  \text{tr}\bigg[[i\gamma\cdot(q + xP)  \nonumber\\
  & \quad - M]\,\sigma_{\mu\nu}\,[i\gamma\cdot(q + xP) - M]\,\gamma_\beta^T\,\nonumber\\
  &\quad\times [i\gamma\cdot(q + (x-1)P) - M]\,\gamma_\alpha^T\bigg]\nonumber\\
  &\quad\times[q^2 - x(1 - x)m_{qq_1}^2 + M^2]^{-3} \\
  &\to 16iM \bar N E_{qq_1}^2\Big(\mathbb P_{\alpha\mu}(P)\mathbb P_{\beta\nu}(P) - \mathbb P_{\alpha\nu}(P)\mathbb P_{\beta\mu}(P)\Big)\nonumber\\
  &\quad \times \int_0^1dx\,(1 - x)\Big\{\Big(M^2 - x(x - 2)m_{qq_1}^2\Big)\,\mathcal G_1^{\rm iu}(\omega) \nonumber\\
  &\quad + \mathcal G_2^{\rm iu}\Big(x(x - 1)m_{qq_1}^2 + M^2\Big)\Big\}
  %\Big|_{\omega = x(x - 1)m_{qq_1}^2 + M^2}\,,
    \label{D4Tensor}
\end{align}}
\hspace*{-0.5\parindent}where $P$ is the incoming and outgoing momentum of the $1^+$ diquark, and $\mathcal G_1^{\rm iu}(\omega)$, $\mathcal G_2^{\rm iu}(\omega)$ are defined in Eqs.\,\eqref{DFG1}, \eqref{DFG2}.  Applying the projector in Eq.\,\eqref{ProjectorTensor} and evaluating the resulting trace, one finds %\begin{subequations}
\begin{equation}
  \delta_{T4} u = \frac{\delta_4 q_0}{2} + \delta_4 q_+ = 0.292\,, \;  %0.150053
  \delta_{T4} d = \frac{\delta_4 q_0}{2} = 0.0589\,.  % 0.0302409
\end{equation}
%\end{subequations}
% 0.292113
% 0.0588755

\subsubsection{Diagram 5 -- tensor}
This is the contribution to the tensor charge arising when a scalar diquark absorbs the tensor probe and is thereby transformed into a $1^+$ diquark.  Naturally, in a symmetry preserving treatment of any reasonable interaction, this contribution is identical to that produced by Diagram~6.  Concretely, one has:
\begin{align}\label{T6}
  &\delta_5 q\,\Lambda^+(p)\sigma_{\mu\nu}\Lambda^+(p) \nonumber\\
  &= {\mathcal N}_5^\kappa\,\Lambda^+(p)\,{\mathcal A}_\alpha^0(-p)\int\frac{d^4\ell}{(2\pi)^4}\,\Delta_{\alpha\beta}^{1^+}(\ell + p)\mathcal V^{10}_{\beta\mu\nu}(\ell + p)\nonumber \\
  &\quad\times \Delta^{0^+}(\ell + p)S(-\ell)\,\mathcal S\,\Lambda^+(p)   \\
%%
%  &= -\mathcal N\,\Lambda^+(p)\,\Big(a_1^0\gamma^5\gamma_\alpha + ia_2^0\gamma^5\hat p_\alpha\Big)\int\frac{d^4\ell}{(2\pi)^4}\frac{\mathbb P_{\alpha\beta}(\ell + p)}{(\ell + p)^2 + m_{qq_1}^2}\,\mathcal V_{\beta\mu\nu}(\ell + p)\,\frac{1}{(\ell + p)^2 + m_{qq_0}^2}\frac{-i\gamma\cdot\ell - M}{\ell^2 + M^2}\,s\,\Lambda^+(p) \nonumber\\
%%
  &= -2\,{\mathcal N}_5^\kappa\,\Lambda^+(p)\,\gamma_5\Big(a_1^0\gamma_\alpha + ia_2^0\hat p_\alpha\Big)\int_0^1 dx\int_0^1 dy\,y\int\frac{d^4\ell}{(2\pi)^4}\nonumber\\
  &\quad\times [i\gamma\cdot(-\ell + yp) - M] \mathcal V^{10}_{\beta\mu\nu}(\ell + (1 - y)p)\,\nonumber\\
  &\quad\times \mathbb P_{\alpha\beta}(\ell + (1 - y)p)\,s\,\Lambda^+(p)
  [\ell^2 + y(y - 1)m_N^2  \nonumber\\
  &\quad + xym_{qq_1}^2 + (1 - x)ym_{qq_0}^2 + (1 - y)M^2]^{-3} \,.
\end{align}

The transition vertex is $\mathcal V^{10}_{\beta\mu\nu}(P,P)$ where ($\bar N = 2$)
\begin{align}
  &\mathcal V^{10}_{\beta\mu\nu}(P,P^\prime)
  = -2\bar N\int\frac{d^4q}{(2\pi)^4}\,\text{tr}\Big\{S(q + P')\sigma_{\mu\nu} S(q + P)\nonumber\\
  &\quad\times \Gamma_{qq}^{0^+}(P)S(q)\bar\Gamma_{qq\beta}^{1^+}(-P')\Big\} \\
%  &= 2N_c\int\frac{d^4q}{(2\pi)^4}\,\text{tr}\bigg\{\frac{i\gamma\cdot(q + P') - M}{(q + P')^2 + M^2}\,\sigma_{\mu\nu}\,\frac{i\gamma\cdot(q + P) - M}{(q + P)^2 + M^2}\,\gamma^5\Big(i E_{qq_0} + \frac{1}{M}\,\gamma\cdot P\,F_{qq_0}\Big)\,\frac{i\gamma\cdot q - M}{q^2 + M^2}\,iE_{qq_1}\gamma_\beta^T(P')\bigg\} \nonumber\\
%%
  &= 4i \bar N E_{qq_1}\int_0^1dx\int_0^1dy\,y\int\frac{d^4q}{(2\pi)^4}\,\nonumber\\
  &\quad \times \text{tr}\Big\{[i\gamma\cdot(q + yP' - xyP) - M]\,\sigma_{\mu\nu}\,\nonumber\\
  &\quad \times [i\gamma\cdot(q - (1 - y)P' + (1 - xy)P) - M] \nonumber\\
  &\quad\times\gamma_5\Big(i E_{qq_0} + \frac{1}{M}\,\gamma\cdot P\,F_{qq_0}\Big)\,\nonumber\\
  &\quad\times [i\gamma\cdot(q - (1 - y)P' - xyP) - M]\,\gamma_\beta^T(P')\Big\} \nonumber\\
  &\quad\times\Big(q^2 - (1 - x)y(1 - y)m_{qq_1}^2 \nonumber\\
  &\quad\quad - x(1 - x)y^2m_{qq_0}^2 + M^2\Big)^{-3},
\end{align}
where $P$ and $P'$ are the incoming and outgoing momenta of the diquarks, respectively. (Some details about the on-shell procedure can be found in App.\,\ref{AppTD}.) Applying the projector in Eq.\,\eqref{ProjectorTensor}, evaluating the resulting trace and combining the result with that from Diagram~6, one finds
\begin{align}
  \delta_{T,{5+6}} u = \delta_{T,{5+6}} d = \delta q_5 = -0.164\,.
\end{align}
%% -0.20997 -> -0.190801
%% -0.16439

\subsubsection{Proton tensor charge}
The results obtained from all diagrams considered in connection with the proton's tensor charges are collected in Table~\ref{TT}.  Notably, the values of the tensor charges depend on the renormalisation scale associated with the tensor vertex.  This is discussed in App.\,\ref{AppEvolution}.

\begin{table}[t!]
\begin{center}
\begin{tabular}{|l|c|c|c|c|}
\hline
 & $\delta_T u$ & $\delta_T d$ &  $g_T^{(0)}$ & $g_T^{(1)}$ \\
\hline\hline
%Diagram 1 & $\phantom{-}0.625$ & $0$ & $\phantom{-}0.625$ & $0.625$ \\
Diagram 1 & $\phantom{-}0.581$ & $0$ & $\phantom{-}0.581$ & $0.581$ \\
\hline
%Diagram 2 & $-0.020$ & $-0.040$ & $-0.060$ & $0.020$ \\
Diagram 2 & $-0.018$ & $-0.036$ & $-0.054$ & $0.018$ \\
\hline
Diagram 3 & $\phantom{-}0$ & $0$ & $\phantom{-}0$ & $0$ \\
\hline
%Diagram 4 & $\phantom{-}0.150$ & $\phantom{-}0.030$ & $\phantom{-}0.180$ & $0.120$ \\
Diagram 4 & $\phantom{-}0.292$ & $\phantom{-}0.059$ & $\phantom{-}0.351$ & $0.233$ \\\hline
Diagram 5+6 & $-0.164$ & $-0.164$ & $-0.329$ & $0$ \\
\hline\hline
%Diagram 6 & $-0.102$ & $-0.102$ & $-0.203$ & $0$ \\
%\hline\hline
%Total Result & $\phantom{-}0.545$ & $-0.219$ & $\phantom{-}0.466$ & $0.814$ \\
Total Result & $\phantom{-}0.691$ & $-0.141$ & $\phantom{-}0.550$ & $0.832$ \\
\hline
%JLab & $\phantom{-}0.54\phantom{0}$ & $-0.23\phantom{0}$ & $\phantom{-}0.31\phantom{0}$ & $0.77\phantom{0}$ \\\hline
\end{tabular}
\end{center}
\caption{Summary of results computed from all diagrams considered in connection with the proton's tensor charge.  They represent values at the model scale, $\zeta_H\approx M$, described in App.\,\ref{AppModelScale}.
\label{TT}}
\end{table}

\subsubsection{Proton tensor charge -- scalar diquark only}
\label{TensorScalarOnly}
It is interesting to consider the impact of the axial-vector diquark on the tensor charges.  This may be exposed by comparing the results in Table~\ref{TT} with those obtained when the axial-vector diquark is eliminated from the nucleon.  We implement that suppression by using the following nucleon Faddeev amplitude:
\begin{equation}
\label{NucleonEigenVectorScalar}
\begin{array}{ccccc}
s(P) & a_1^+(P) & a_1^0(P) & a_2^+(P) & a_2^0(P)\\
1.0 & 0 & 0 & 0  & 0
\end{array}\,,
\end{equation}
and then repeating the computations in Apps.\,\ref{emApp}, \ref{AppTensor}.  Naturally, in this case only Diagrams~1 and 3 can possibly yield nonzero contributions to any quantity.

Recomputing the canonical normalisation, we obtain
\begin{equation}
\mathcal N_{\not \,1} = \frac{1}{0.0174} = 57.50\,,
\end{equation}
%\frac{1}{0.0247662} = 40.3776
%1/0.019864
which is 2\% larger than the complete result in Eq.\,\eqref{normalisation}.

Regarding the tensor charges, Diagram~3 also vanishes in this instance so that the net result is simply that produced by Diagram~1:
\begin{equation}
\label{NoAxial}
\begin{array}{cccc}
\delta_{T\!\not\, 1} u & \delta_{T\!\not\, 1} d &  g_{T\!\not\, 1}^{(0)} & g_{T\!\not\, 1}^{(1)} \\
0.765 & 0 & 0.765 & 0.765
\end{array}\,.
\end{equation}
%% Mario says revised value is 0.765374.
Comparison with Table~\ref{TT} shows that with a symmetry-preserving treatment of a vector$\,\otimes\,$vector contact interaction, the $d$-quark contribution to the proton's tensor charge is only nonzero in the presence of axial-vector diquark correlations and these correlations reduce the $u$-quark contribution by 10\%.

\subsubsection{Proton tensor charge -- Reduced DCSB}
\label{TensorDCSBCut}
In order to expose the effect of DCSB on the tensor charges, we repeated all relevant calculations above beginning with the value of $\alpha_{\rm IR}$ used to produce Eq.\,\eqref{dressedMvalue08} and thereby obtained the results listed in Table~\ref{TT08}.

\begin{table}[t!]
\begin{center}
\begin{tabular}{|l|c|c|c|c|}
\hline
 & $\delta_T u$ & $\delta_T d$ &  $g_T^{(0)}$ & $g_T^{(1)}$ \\
\hline\hline
Diagram 1 & $\phantom{-}0.495$ & $0$ & $\phantom{-}0.495$ & $0.495$ \\
\hline
Diagram 2 & $-0.020$ & $-0.039$ & $-0.059$ & $0.020$ \\
\hline
Diagram 3 & $\phantom{-}0$ & $0$ & $\phantom{-}0$ & $0$ \\
\hline
Diagram 4 & $\phantom{-}0.236$ & $\phantom{-}0.047$ & $\phantom{-}0.283$ & $0.189$ \\
\hline
Diagram 5+6 & $-0.160$ & $-0.160$ & $-0.319$ & $0$ \\
\hline\hline
%Diagram 6 & $-0.102$ & $-0.102$ & $-0.203$ & $0$ \\
%\hline\hline
Total Result & $\phantom{-}0.551$ & $-0.151$ & $\phantom{-}0.400$ & $0.703$ \\
\hline
%JLab & $\phantom{-}0.54\phantom{0}$ & $-0.23\phantom{0}$ & $\phantom{-}0.31\phantom{0}$ & $0.77\phantom{0}$ \\\hline
\end{tabular}
\end{center}
\caption{Summary of results computed from all diagrams considered in connection with the proton's tensor charge using input based on $\alpha_{\rm IR}/\pi = 0.74$, quoted at the model scale, $\zeta_H\approx M$, described in App.\,\ref{AppModelScale}.
\label{TT08}}
\end{table}

\section{On-shell Considerations for the Transition Diagrams}
\label{AppTD}
For the practitioner it will likely be helpful here to describe our treatment of the denominator that arises when using a Feynman parametrisation to compute the transition diagrams.  Namely, one has
\begin{align}
  &\frac{1}{(q + P')^2 + M^2}\frac{1}{(q + P)^2 + M^2}\frac{1}{q^2 + M^2} \nonumber\\
%  &=2\int_0^1dx\int_0^1dy\,y
%  [\{(1 - y)(q + P')^2 \nonumber\\
%  & \quad + xy(q + P)^2 + (1 - x)yq^2 + M^2\}^{-3} \nonumber\\
  &=2\int_0^1dx\int_0^1dy\,y \{(q + (1 - y)P' + xyP)^2 \nonumber \\
  & \quad + (1 - y)yP'^2 + xy(1 - xy)P^2 \nonumber \\\
  & \quad - 2(1 - y)xyP'\cdot P + M^2\}^{-3}\,.
\end{align}
At this point, a shift of the integration variable $q \to q - (1 - y)P' - xyP$ yields
\begin{align}
  & 2\int_0^1dx\int_0^1dy\,y\{q^2 + (1 - y)yP'^2 + xy(1 - xy)P^2 \nonumber\\
  & \quad - 2(1 - y)xyP'\cdot P + M^2\}^{-3}\,.
\end{align}

Next, we employ on-shell relations, which for Diagram~5 are given by
\begin{equation}
  P'^2 = -m_{qq_1}^2\,, \;
  P^2 = -m_{qq_0}^2\,.
\end{equation}
Then, since $Q^2 \equiv (P' - P)^2 = P'^2 + P^2 - 2P'\cdot P = 0$:
\begin{equation}
  P'\cdot P = -\frac{m_{qq_0}^2 + m_{qq_1}^2}{2}\,.
\end{equation}
Hence, the Feynman integral associated with Diagram~5 is
\begin{align}
  & 2\int_0^1dx\int_0^1dy\,y \{q^2 - (1 - x)y(1 - y)m_{qq_1}^2 \nonumber \\
  & \quad - x(1 - x)y^2m_{qq_0}^2 + M^2\}^{-3}\,.
\end{align}
Diagram~6 is obtained via $m_{qq_0} \leftrightarrow m_{qq_1}$.

\section{Model Scale}
\label{AppModelScale}
In modern studies of QCD's gap equation, which use DCSB-improved kernels and interactions that preserve the one-loop renormalisation group behaviour of QCD, the dressed-quark mass is renormalisation point invariant.  As in QCD, however, the current-quark mass is not.  Therefore, in quoting a current-quark mass in Eq.\,\eqref{parametervalues}, a question immediately arises: to which scale, $\zeta_H$, does this current-quark mass correspond?

As noted in App.\,\ref{sec:contact}, the contact-interaction does not define a renormalisable theory and the scale $\zeta_H$ should therefore be part of the definition of the interaction.  We define $\zeta_H$ so as to establish contact between the current-quark mass in Eq.\,\eqref{parametervalues} and QCD.

Current-quark masses in QCD are typically quoted at a scale of $\zeta_2 = 2\,$GeV.  A survey of available estimates indicates \cite{Beringer:1900zz}
\begin{equation}
m(\zeta_2) = \frac{m_u(\zeta_2) + m_d(\zeta_2)}{2} = 3.5^{+0.7}_{-0.2}\,;
\end{equation}
and this value compares well with that determined from a compilation of estimates using numerical simulations of lattice-regularised QCD \cite{Colangelo:2010et}:
\begin{equation}
m(\zeta_2) = \frac{m_u(\zeta_2) + m_d(\zeta_2)}{2} = 3.4 \pm 0.2\,.
\end{equation}
On the other hand, we have determined an average value of the $u$- and $d$-quark masses appropriate to our interaction that is $m(\zeta_H):=m= 7\,$MeV.

The scale dependence of current-quark masses in QCD is expressed via
\begin{equation}
\frac{m(\zeta^\prime)}{m(\zeta)} = \left[\frac{\alpha_s(\zeta^\prime)}{\alpha_s(\zeta)}\right]^{\gamma_m}\,,
% log[zeta] / log[zeta-prime] = al(z-prime)/al(z)
\end{equation}
where $\alpha_s(\zeta)$ is the running coupling and $\gamma_m=12/(33-2 n_f)$, with $n_f$ the number of active fermion flavours, is the mass anomalous dimension.  Plainly, the running current-quark mass increases as the scale is decreased.

Using the one-loop running coupling,
with $n_f=4$ and $\Lambda_{\rm QCD}=0.234\,$GeV \cite{Qin:2011dd}, then
\begin{equation}
\label{ModelScale}
m(\zeta_H) \approx 2  m(\zeta_2) \quad \mbox{for} \quad \zeta_H = 0.39 \pm 0.02\,{\rm GeV}\,;
\end{equation}
and thus we have determined the model-scale.  Given the arguments in Refs.\,\cite{Holt:2010vj,Chang:2012rk,Chang:2014lva}, the outcome $\zeta_H \approx M$ is both internally consistent and reasonable.  (We use the one-loop expression owing to the simplicity of our framework.  Employing next-to-leading-order (NLO) evolution leads simply to a 25\% increase in $\zeta_{\rm H}$ with no material phenomenological differences.)

\section{Scale Dependence of the Tensor Charge}
\label{AppEvolution}
Whilst the values of the tensor charges are gauge- and Poincar\'e-invariant, they depend on the renormalisation scale, $\zeta$, employed to compute the dressed inhomogeneous tensor vertex
\begin{equation}
\Gamma_{\mu\nu}(k;Q;\zeta)=S_1(k;Q;\zeta)\sigma_{\mu\nu}+\ldots\,,
\end{equation}
at zero total momentum, $Q=0$.  ($k$ is the relative momentum.)  The renormalisation constant $Z_T(\zeta,\Lambda)$ is the factor required as a multiplier for the Bethe-Salpeter equation inhomogeneity, $\sigma_{\mu\nu}$, in order to achieve $S_1(k^2=\zeta^2;Q=0;\zeta) = 1$.

At one-loop order in QCD \cite{Barone:1997fh}:
\begin{equation}
\label{GammaTzeta}
\Gamma_{\mu\nu}(k;Q;\zeta) \stackrel{\zeta^2 \gg \Lambda_{\rm QCD}^2}{=} \left[\frac{\alpha_S(\zeta_0^2)}{\alpha_S(\zeta^2)}\right]^{\eta_T}
\Gamma_{\mu\nu}(k;Q;\zeta_0)\,,
\end{equation}
where $\eta_T=(-1/3) \gamma_m$.  The pointwise behaviour of $\Gamma_{\mu\nu}(k;Q=0;\zeta)$ is illustrated in Ref.\,\cite{Yamanaka:2013zoa}.
%%%
%We note that
%\begin{equation}
%S_1(k^2;P=0;\zeta) \stackrel{k^2\gg\Lambda_{\rm QCD}^2}{=}
%\left[\frac{\alpha_S(\zeta^2)}{\alpha_S(k^2)}\right]^{\eta_T}\,.
%\end{equation}
%%%fit seems to suggest -3/25 instead of -4/25.

%Then, for all k^2>>LQCD^2
%    S1(k^2,zeta^2) = ( Log[zeta/LQCD] / Log[Sqrt[k^2]/LQCD] )^(4/25)
%During evaluation of In[45]:= rescaling factor 2/zetaH
%Out[58]= 0.793701
%Out[59]= 0.4131
%Out[60]= 0.808525
%Out[61]= 0.3662
%Out[62]= 0.778279
%% 0.794 +
Equation~\eqref{GammaTzeta} entails
\begin{equation}
\delta q(\zeta) \stackrel{\zeta^2 \gg \Lambda_{\rm QCD}^2}{=} \left[\frac{\alpha_S(\zeta_0^2)}{\alpha_S(\zeta^2)}\right]^{\eta_T} \delta q(\zeta_0)\,,\rule{0ex}{4ex}
\end{equation}
and hence that $\delta q$ decreases as $\zeta$ increases.  It follows, for example and in connection with our analysis, that
\begin{equation}
\label{Efactor}
\frac{\delta q(\zeta_2)}{\delta q(\zeta_H)} = 0.794 \pm 0.015\,,
\end{equation}
with $\zeta_H$ drawn from Eq.\,\eqref{ModelScale}.  %N.B.\,Any significant differences generated by next-to-leading-order evolution are masked by a 25\% increase in $\zeta_{\rm H}$ \cite{Gluck:1999xe} and hence are immaterial herein.

\section{Euclidean Conventions}
\label{App:EM}
In our Euclidean formulation:
\begin{equation}
p\cdot q=\sum_{i=1}^4 p_i q_i\,;
\end{equation}
% \begin{equation}
\begin{eqnarray}
&& \{\gamma_\mu,\gamma_\nu\}=2\,\delta_{\mu\nu}\,;\;
\gamma_\mu^\dagger = \gamma_\mu\,;\;
\sigma_{\mu\nu}= \frac{i}{2}[\gamma_\mu,\gamma_\nu]\,; \rule{2em}{0ex}\\
&& {\rm tr}\,[\gamma_5\gamma_\mu\gamma_\nu\gamma_\rho\gamma_\sigma] =
-4\,\epsilon_{\mu\nu\rho\sigma}\,, \epsilon_{1234}= 1\,.
\end{eqnarray}

A positive energy spinor satisfies
\begin{equation}
\label{DiracEqn}
\bar {\mathpzc u}(P,s)\, (i \gamma\cdot P + M) = 0 = (i\gamma\cdot P + M)\, {\mathpzc u}(P,s)\,,
\end{equation}
where $s=\pm \frac{1}{2}$ is the spin label.  The spinor is normalised:
\begin{equation}
\bar {\mathpzc u}(P,s) \, {\mathpzc u}(P,s) = 2 M \,,
\end{equation}
and may be expressed explicitly:
\begin{equation}
{\mathpzc u}(P,s) = \sqrt{M- i {\cal E}}
\left(
\begin{array}{l}
\chi_s\\
\displaystyle \frac{\vec{\sigma}\cdot \vec{P}}{M - i {\cal E}} \chi_s
\end{array}
\right)\,,
\end{equation}
with ${\cal E} = i \sqrt{\vec{P}^2 + M^2}$,
\begin{equation}
\chi_+ = \left( \begin{array}{c} 1 \\ 0  \end{array}\right)\,,\;
\chi_- = \left( \begin{array}{c} 0\\ 1  \end{array}\right)\,.
\end{equation}
For the free-particle spinor, $\bar {\mathpzc u}(P,s)= {\mathpzc u}(P,s)^\dagger \gamma_4$.

The spinor can be used to construct a positive energy projection operator:
\begin{equation}
\label{Lplus} \Lambda_+(P):= \frac{1}{2 M}\,\sum_{s=\pm} \, {\mathpzc u}(P,s) \, \bar
{\mathpzc u}(P,s) = \frac{1}{2M} \left( -i \gamma\cdot P + M\right).
\end{equation}

A charge-conjugated Bethe-Salpeter amplitude is obtained via
\begin{equation}
\label{chargec}
\bar\Gamma(k;P) = C^\dagger \, \Gamma(-k;P)^{\rm T}\,C\,,
\end{equation}
where ``T'' denotes a transposing of all matrix indices and
$C=\gamma_2\gamma_4$ is the charge conjugation matrix, $C^\dagger=-C$.  We note that
\begin{equation}
C^\dagger \gamma_\mu^{\rm T} \, C = - \gamma_\mu\,, \; [C,\gamma_5] = 0\,.
\end{equation}

%\bibliographystyle{../../../../zProc/z10KITPC/h-physrev4}
%\bibliography{../../../../CollectedBiB}

\end{document}